\documentclass[12pt]{article}
\usepackage{cite}
\usepackage{epsfig}
\usepackage{graphicx}
\usepackage{amsmath}
\usepackage{amssymb}
\usepackage{ulem}
\usepackage{mdwlist} 
\usepackage{color}  
\usepackage[OT2, T1]{fontenc}
\usepackage[english]{babel}
\usepackage{pifont}
\usepackage{hyperref}

\usepackage{a41}
\usepackage{color}
\usepackage[rflt]{floatflt}
\usepackage{float}
\usepackage{slashed}
\usepackage{multirow}

\def\q{\slashed{q}} 

\usepackage{array}

\usepackage[english]{babel}

\usepackage{url}

\usepackage{amsmath, amsthm, amssymb}
\newtheorem{thm}{Theorem}[section]

\def\b0{\beta_0}

\newtheorem{definition}[thm]{Definition}

 \newcommand{\LL}{\ln\left[\frac{1-x}{1+x}\right]}

\newcommand{\rr}{[s - 16]}

\newcommand{\sign}[1]{\textnormal{sign}\hspace{-0.1em}\left(#1\right)}

\newcommand{\Li}{{\rm Li}}
\newcommand{\HA}{{\rm H}}

\newcommand{\ep}{\varepsilon}

\def\b0{\beta_0}

\usepackage{rotating}

\usepackage{graphicx}

\newcounter{mmacnt}
\def\restartmma{\setcounter{mmacnt}{0}}
\restartmma \catcode`|=\active
\def|#1|{\mathrm{#1}}
\catcode`|=12
\newenvironment{mma}{
 \par\smallskip
 \catcode`|=\active
 \parskip=0pt\parindent=0pt 
 \small
 \def\In##1\\{%
\def\linebreak{\hfill\break\null\qquad}%
\refstepcounter{mmacnt}
\hangindent=2.5em\hangafter=0
\leavevmode
\llap{\tiny\sffamily n[\arabic{mmacnt}]:=\kern.5em}%
\mathversion{bold}\footnotesize$\displaystyle##1$\normalsize
\mathversion{normal}\par
 }%
 \def\Print##1\\{%
\def\linebreak{\hfill\break}%
\hangindent=2.5em\hangafter=0
\leavevmode ##1\par}%
 \def\Out##1\\{%
\def\linebreak{$\hfill\break\null\hfill$}%
\kern\abovedisplayskip\par
\hangindent=2.5em\hangafter=0
\leavevmode
\llap{\tiny\sffamily Out[\arabic{mmacnt}]=\kern.5em}
\footnotesize$\displaystyle##1$\normalsize\hfill\null\par
\kern\belowdisplayskip
 }%
 \def\Warning##1##2\\{%
\def\linebreak{\hfill\break}%
\hangindent=2.5em\hangafter=0
\leavevmode
{\scriptsize##1 : ##2}\par}%
}{%
 \par\smallskip
}

\usepackage{color}

\newenvironment{fshaded}{%
\MakeFramed {\FrameRestore}
}%
{\endMakeFramed}

\catcode`,\active

\catcode`\,12

\allowdisplaybreaks[4]

\begin{document}
\setlength{\baselineskip}{0.515cm}
\sloppy
\thispagestyle{empty}
\begin{flushleft}
DESY 26--118 \\
RISC Report number 26--12\\
September 2026\\
\end{flushleft}

\mbox{}
\vspace*{\fill}
\begin{center}

{\LARGE\bf  }

\vspace*{2mm}
{\LARGE\bf  Analytic results on the massive three-loop} 

\vspace*{3mm}
{\LARGE\bf form factors: gluonic contributions}

\vspace{3cm}
\large 
J.~Bl\"umlein$^{a,b}$,
A.~De~Freitas$^c$,
P.~Marquard$^a$, 
J.~Obrovsky$^c$,
and C.~Schneider$^c$

\vspace{1.cm}
\normalsize
{\it  $^a$Deutsches Elektronen--Synchrotron DESY,
Platanenallee 6, 15738 Zeuthen, Germany}

\vspace*{2mm}
{\it $^b$Institut f\"ur Theoretische Physik III, IV, TU Dortmund, \\ Otto-Hahn
Stra\ss{}e 4, 44227 Dortmund, Germany}

\vspace*{2mm}
{\it $^c$Johannes Kepler University Linz,
Research Institute for Symbolic Computation (RISC), Altenbergerstra{\ss}e 69,
A--4040, Linz, Austria}


\end{center}
\normalsize
\vspace{\fill}
\begin{abstract}
\noindent
We compute the gluonic contributions to the three-loop heavy-quark form factors for the 
vector, axial-vector, scalar, and pseudoscalar currents. In the low-energy limit, $q^2/m^2 
\rightarrow 0$, we used guessing algorithms to derive closed-form difference and differential 
equations from the rational sequences associated with the multiple zeta values and other 
constants appearing in the expansion coefficients, which required up to 26000 coefficients 
in the most demanding cases. Part of the results are obtained analytically in terms of 
harmonic polylogarithms and square-root valued iterated integrals by solving the obtained 
differential equations. For the remaining contributions, arbitrarily deep series expansions 
around the singularities of the form factors can be obtained by matching local expansions at 
intermediate points and by exploiting the differential equations obeyed by the functions 
associated with each transcendental constant in the expansions around $s = q^2/m^2 = 0$. 
For all these calculations advanced computer algebra methods have been employed. By using 
analytic continuation methods for differential equations matching the expansions at different 
values of $s$ and using PSLQ, we derive analytic results in the high-energy limit, $q^2/m^2 
\rightarrow \infty$. The expansion coefficients are expressed in terms of multiple zeta 
values up to weight {\sf w}~=~6 together with three additional constants, two related to 
sixth-root-of-unity letters and one associated with quadratic-form implied iterated 
integrals. We also derive deep expansions about the threshold and pseudo-threshold. Numerical 
results are presented in the whole kinematic range and compared to results in the literature.
\end{abstract}

\vspace*{\fill}

\numberwithin{equation}{section}
\newpage 
\section{Introduction} 
\label{sec:1}

\vspace*{1mm}
\noindent
The massive form factors associated with the vector, axial-vector, scalar, and pseudoscalar currents describe 
the coupling of virtual bosons to a pair of massive quarks in Quantum Chromodynamics (QCD) and Quantum 
Electrodynamics (QED). They constitute fundamental building blocks for precision calculations involving heavy 
quarks and enter a wide range of observables relevant to collider phenomenology, including heavy-quark 
production and Higgs boson decays. The high experimental precision achieved in measurements of these 
processes requires equally precise theoretical predictions, motivating the computation of the corresponding 
form factors at increasingly high loop orders.

Beyond their phenomenological relevance, massive form factors provide one of the simplest settings for 
investigating the infrared structure of gauge theories with massive particles. They therefore play a 
central role in the study of infrared singularities and serve as benchmark quantities for the development 
of modern methods for multi-loop calculations.

The evaluation of the three-loop massive form factors represents a particularly challenging problem due to 
the large number of contributing Feynman diagrams and the complexity of the associated Feynman integrals, 
which involve structures beyond standard harmonic polylogarithms and related 
functions.\footnote{For a survey see Ref.~\cite{Blumlein:2018cms}.} Considerable 
progress has been achieved in recent years. Partial analytic results for the three-loop form 
factors have been obtained in 
Refs.~\cite{Gluza:2009yy,Henn:2016kjz,Henn:2016tyf,Ahmed:2017gyt,Ablinger:2017hst,
Ablinger:2018yae,Ablinger:2018zwz,Lee:2018nxa,Lee:2018rgs,Blumlein:2018tmz,Blumlein:2019oas,Blumlein:2023uuq},
while complete numerical results at this order were presented in Refs.~\cite{Fael:2022miw,Egner:2022jot,
Fael:2022rgm,Schonwald:2022djs,Fael:2023zqr,Schonwald:2023uel}.\footnote{Complete analytic results are known 
at $s=0$ in the form of series expansions.}

In this paper we complete the analytic determination of the three-loop QCD form factors by 
computing the purely gluonic contributions, thereby complementing our previous calculation 
of the quarkonic contributions presented in Ref.~\cite{Blumlein:2023uuq}.\footnote{For QED 
the QCD color factors are replaced according to $C_F=(N_c^2-1)/(2N_c)\rightarrow1$, 
$T_F=1/2\rightarrow1$, and $C_A=N_c\rightarrow0$, with $N_c=3$ in QCD.} 
While the planar color contributions were obtained previously in Ref.~\cite{Ablinger:2018zwz},
the present work provides the first complete analytic treatment including all non-planar 
diagrams. The vector, axial-vector, scalar and pseudoscalar form factors are given by
\begin{equation}
\label{eq:def}
F_I(s) = 1 + \sum_{k=1}^\infty a_s^k(m^2) F_I^{(k)}(s),
\end{equation}
where 
\begin{equation} 
\label{eq:kin} 
s =: \frac{q^2}{m^2}, 
\end{equation} 
and $q^2$ denotes the virtuality of the incoming current, $m$ the heavy-quark 
mass, and $a_s = \alpha_s/(4\pi)$ is the running coupling constant 
renormalized in the $\overline{\rm MS}$ scheme, cf.~Refs.~\cite{Tarasov:1980au,
Larin:1993tp,vanRitbergen:1997va,Czakon:2004bu,Chetyrkin:2004mf,
Baikov:2016tgj,Herzog:2017ohr,Chetyrkin:2017bjc,Luthe:2017ttg}. 
For the analytic representation of the form factors, it is convenient to introduce the 
variable $x$, related to the dimensionless kinematic variable $s$ by
\begin{equation} 
\label{eq:x2s} 
s=-\frac{(1-x)^2}{x}~~~~~\text{and}~~~~~x = 
\frac{\sqrt{4-s}-\sqrt{-s}}{\sqrt{4-s}+\sqrt{-s}}. 
\end{equation} 
Our approach relies as far as possible on analytic and symbolic methods. As in many multi-loop calculations, 
the amplitudes are first reduced to a set of master integrals, which satisfy coupled systems of differential 
equations. The main methodological advance of this work is the systematic exploitation of these equations to
generate high-order expansions in the low energy limit. The resulting expansion coefficients are then used 
to derive recursion relations and differential equations from the sequences of rational numbers associated 
with the transcendental constants appearing in the expansions. This is done first at the level of the master 
integrals, and later at the level of the full physical expressions for the form factors. The most 
challenging differential equations do not factorize into first-order operators. In these cases we employ 
Frobenius' method \cite{FROBENIUS}, constructing high-order local series expansions 
\cite{COHI,HOEVEN,MEZZA,Grigo:2012ji,ABRAMOV1} around regular and singular 
points and matching overlapping expansions to obtain analytic continuations throughout the entire kinematic 
range.

The form factors have a threshold at $s = 4$ and a pseudo-threshold at $s = 16$.
A main result of the present paper is the analytic result for the massive 
three-loop form factors in the high energy limit $x \rightarrow 0$ or $s 
\rightarrow \infty$, computed for the first time. At  other characteristic 
points, we provide deep series expansions. Furthermore, we provide semi-analytic 
series representations at a relative accuracy of $O(10^{-30})$ in {\tt Mathematica} 
format and associated numerical implementations in the whole kinematic domain.
  
The paper is organized as follows. In Section~\ref{sec:2}, we describe the 
structure of vector, axial-vector, scalar, and pseudoscalar currents in terms 
of form factors and outline the basic calculation strategies. While these 
strategies have already been employed by us to calculate the quarkonic parts of 
the three-loop form factors in Refs.~\cite{Blumlein:2019oas,Blumlein:2023uuq}, 
the more challenging gluonic case required substantial refinements. These 
improvements are carried out in the three following sections. 
Section~\ref{Sec:LargeMoments} derives the coefficients for the series expansions 
of the form factors at $s=0$ ($x=1$). Section~\ref{Sec:Guessing} presents 
the guessing of their underlying recurrences and differential equations, 
and Section~\ref{Sec:AnalyticCont} derives expansions at other relevant expansion 
points using efficient methods of analytic continuation. Based on these 
refinements, in Section~\ref{Sec:HighEnergyLimit} we describe how the analytic 
high energy result in the limit $s \rightarrow \infty$ is obtained. Series 
expansions around $s = 4$ and $s=16$ are discussed in 
Section~\ref{Sec:NumericalResults} and the conclusions are given in 
Section~\ref{Sec:Conclusion}. The ancillary files to the present paper contain computer-readable 
analytic codes. They are described in Appendix~\ref{sec:A}.
\section{The basic calculation strategy}
\label{sec:2}

\begin{center}
\begin{figure}[h]
\begin{center}
\begin{minipage}[c]{0.15\linewidth}
     \includegraphics[width=1\textwidth]{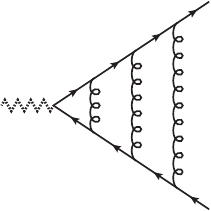}
\vspace*{-10mm}
\begin{center}
{\footnotesize{(a)}}
\end{center}
\end{minipage}
\hspace*{0.01\linewidth}
\begin{minipage}[c]{0.15\linewidth}
     \includegraphics[width=1\textwidth]{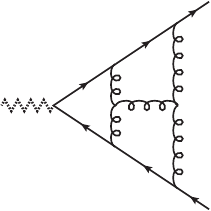}
\vspace*{-10mm}
\begin{center}
{\footnotesize (b)}
\end{center}
\end{minipage}
\hspace*{0.01\linewidth}
\begin{minipage}[c]{0.15\linewidth}
     \includegraphics[width=1\textwidth]{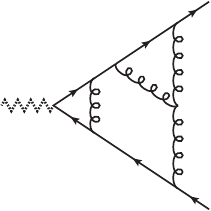}
\vspace*{-10mm}
\begin{center}
{\footnotesize (c)}
\end{center}
\end{minipage}
\begin{minipage}[c]{0.15\linewidth}
     \includegraphics[width=1\textwidth]{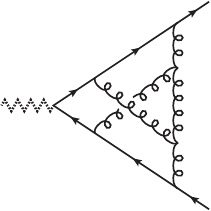}   
\vspace*{-10mm}
\begin{center}
{\footnotesize (d)}
\end{center}
\end{minipage}
\hspace*{0.01\linewidth}
\begin{minipage}[c]{0.15\linewidth}
     \includegraphics[width=1\textwidth]{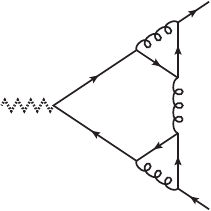}   
\vspace*{-10mm}
\begin{center}
{\footnotesize (e)}
\end{center}
\end{minipage}
\end{center}
\caption{\sf Sample of diagrams required for the calculation of three-loop massive 
form factors. The solid arrow lines represent massive quarks,
while the dotted zigzag line represents any of the possible currents, and the curly lines
gluons. 
}
\label{fig:samplediagrams}
\end{figure}
\end{center}

\vspace*{1mm}
\noindent
The gluonic contributions to the form factors arise exclusively from diagrams in which only 
gluons are exchanged between the external heavy quarks. This is in contrast to the 
considerably 
simpler quarkonic contributions studied in Refs.~\cite{Blumlein:2019oas,Blumlein:2023uuq},
which originate from diagrams containing internal massless and massive quark loops. 
Representative three-loop gluonic diagrams are shown in Figure~\ref{fig:samplediagrams}.
The Feynman diagrams are generated by {\tt QGRAF} \cite{Nogueira:1991ex}, the 
Lorentz- and Dirac-algebra
is carried out in $D = 4 - 2 \ep$ dimensions using {\tt Form} 
\cite{Vermaseren:2000nd,Tentyukov:2007mu} and the color algebra is performed by {\tt 
Color} \cite{vanRitbergen:1998pn}. 

For the description of the different form factors we use the following conventions, see 
also Ref.~\cite{Blumlein:2023uuq}.
Depending on the nature of the boson coupling to the heavy-quark pair, there are four types of 
three-point amplitudes. For a spin-1 vector boson, the vertex function is
\begin{equation}
\Gamma_{V,cd}^{\mu} = -i v_Q \delta_{cd} \left( \gamma^{\mu}~F_{V,1}(s) + \frac{i}{2 m} \sigma^{\mu\nu} q_{\nu} ~F_{V,2}(s) \right)\, ,
\end{equation}
where $\sigma^{\mu\nu}=\frac{i}{2}[\gamma^\mu,\gamma^\nu]$, $c$ and $d$ are the quark color indices,
and $v_Q$ is the corresponding Standard Model vector coupling. The axial-vector vertex is given by
\begin{equation}
\Gamma_{A,cd}^{\mu} = -i a_Q \delta_{cd} \left( \gamma^{\mu}\gamma_5~F_{A,1}(s) + \frac{1}{2m}q^\mu\gamma_5~F_{A,2}(s) \right)\, ,
\end{equation}
where $a_Q$ denotes the Standard Model axial-vector coupling. For a spin-0 scalar boson,
\begin{equation}
\Gamma_{S,cd} = -\frac{m}{v}s_Q\delta_{cd}F_S(s)\, ,
\end{equation}
$v$ is the Higgs vacuum expectation value and $s_Q$ is the scalar coupling. 
Finally, the pseudoscalar vertex is given by
\begin{equation}
\Gamma_{P,cd} = -i\frac{m}{v}p_Q\delta_{cd}\gamma_5F_P(s)\, ,
\end{equation}
and $p_Q$ is the pseudoscalar coupling.

The vector and axial-vector form factors are extracted by applying the projectors
\begin{eqnarray}
P_{V,i} &=& \frac{i}{v_Q}\frac{\delta_{cd}}{N_c}\frac{\q_2-m}{m} \Bigl(\gamma_\mu g_{V,i}^{(1)} +\frac{1}{2m}(q_{2\mu}-q_{1\mu})g_{V,i}^{(2)}\Bigr) \frac{\q_1+m}{m}\,, \\
P_{A,i} &=& \frac{i}{a_Q}\frac{\delta_{cd}}{N_c}\frac{\q_2-m}{m} \Bigl(\gamma_\mu\gamma_5 g_{A,i}^{(1)} +\frac{1}{2m}(q_{1\mu}+q_{2\mu})\gamma_5 g_{A,i}^{(2)}\Bigr) \frac{\q_1+m}{m}\,,
\end{eqnarray}
followed by traces over the spinor and color indices. Here $q_1$ and $q_2$ denote the heavy-quark momenta, satisfying $q=q_1+q_2$.
The coefficients $g_{I,i}^{(k)}$ are given by
\begin{align}
g_{V,1}^{(1)} &=\frac{1}{4 (1-\ep) (4-s)}  \,,  & g_{V,1}^{(2)} &=\frac{3 - 2 \ep}{(1-\ep) (s-4)^2}  \,, \label{eq:gV1} \\
g_{V,2}^{(1)} &=-\frac{1}{(1-\ep) (4-s) s}  \,, & g_{V,2}^{(2)} &=-\frac{2 ((1-\ep) s+2)}{(1-\ep) (4-s)^2 s}  \,, \\
g_{A,1}^{(1)} &=\frac{1}{4 (1-\ep) (4-s)}  \,, & g_{A,1}^{(2)} &=-\frac{1}{(1-\ep) (4-s) s}  \,, \\
g_{A,2}^{(1)} &=-\frac{1}{(1-\ep) (4-s) s}  \,, & g_{A,2}^{(2)} 
&=\frac{2 (\ep (s-4)-s+6)}{(1-\ep) (4-s) s^2} \,. \label{eq:gA2}
\end{align}

For later convenience, we also introduce the individual contributions multiplying the functions $g_{I,i}^{(k)}$ after projection. That is, we decompose the projected form factors as
\begin{eqnarray}
  F_{V,i}(s) &=& g_{V,i}^{(1)}F_{v,1}(s)+g_{V,i}^{(2)}F_{v,2}(s)\, , \qquad i=1,2, \label{Vgidecomp1} \\
  F_{A,i}(s) &=& g_{A,i}^{(1)}F_{a,1}(s)+g_{A,i}^{(2)}F_{a,2}(s)\, , \qquad i=1,2. \label{Agidecomp1}
\end{eqnarray}

The scalar and pseudoscalar form factors are obtained analogously by applying the projectors
\begin{eqnarray}
  P_{S} &=& -\frac{v}{2ms_Q}\frac{\delta_{cd}}{N_c (4-s)} \frac{\q_2-m}{m}\frac{\q_1+m}{m}\,, \\
  P_{P} &=& -\frac{v}{2mp_Q}\frac{\delta_{cd}}{N_c s} \frac{\q_2-m}{m}i\gamma_5\frac{\q_1+m}{m}\,,
\end{eqnarray}
and taking the corresponding spinor and color traces.

In a similar way as in Eqs. (\ref{Vgidecomp1}--\ref{Agidecomp1}),
we extract the $s$-dependent part of the projectors by defining $F_s(s)$ and $F_p(s)$ 
as follows
\begin{eqnarray}
  F_S(s) &=& \frac{1}{2 (s-4)} F_s(s) \, , \label{Sdecomp1} \\
  F_P(s) &=& -\frac{1}{2 s} F_p(s) \, . \label{Pdecomp1}
\end{eqnarray}  

In the following we only consider the three-loop contributions.
In QCD, these form factors are decomposed in terms of the $\mathrm{SU}(N_c)$ color factors as follows
\begin{equation}
  F_I(s) = \sum_{j=1}^3 C_j F_{I;j}(s),
\label{eq:F-color-decomp}
\end{equation}
where the color factors $C_j$ are given by
\begin{equation}
  C_1 = C_F^3\,, \,\,\,\,\,\, C_2 = C_A C_F^2\,, \,\,\,\,\,\, C_3 = C_A^2 C_F.
\label{eq:color-factors}
\end{equation}
To simplify the notation, we henceforth use a single composite index $I$. Depending on the context, $I$ denotes either one of the scalar and pseudoscalar form factors,
$I \in \{S,P\}$, or one of the vector and axial-vector form factors, $I \in \{(V,1),(V,2),(A,1),(A,2)\}$.
Likewise, when referring to the projected components appearing in Eqs.~(\ref{Vgidecomp1}--\ref{Agidecomp1}) and Eqs. (\ref{Sdecomp1}--\ref{Pdecomp1}),
we use $I \in \{(v,1)\, , \, (v,2)\, , \, (a,1)\, , \, (a,2)\, , \, s\, , \,p\}$.

The renormalization of the form factors has been described in detail in Ref.~\cite{Blumlein:2023uuq}
before. The mass and the wave function of the heavy quark have been renormalized using the on-shell 
scheme. For the Yukawa coupling in the scalar and pseudoscalar form factors, 
we also used the 
$\overline{\rm MS}$ renormalization scheme. All the necessary 
renormalization constants are available, 
including the description of the infrared singularities, 
see Refs.~\cite{Schroder:2005hy,Chetyrkin:2005ia,
Chetyrkin:1999ys,Chetyrkin:1999qi,Melnikov:2000qh,Broadhurst:1991fy,Marquard:2018rwx,
Marquard:2016dcn,Marquard:2015qpa,Becher:2009kw,Mitov:2006xs,
Korchemsky:1987wg,Kidonakis:2009ev,Grozin:2014hna,Grozin:2015kna}.

The Feynman amplitudes
are reduced to master integrals (MIs) by applying integration-by-parts 
(IBP) reduction 
\cite{IBP1,IBP2,IBP3,IBP4,Chetyrkin:1981qh,Laporta:2001dd}, using the package 
{\tt Crusher}~\cite{CRUSHER}.
The master integrals admit Laurent expansions in $\ep$ containing pole terms. One forms differential equations
for the master integrals. By a formal Taylor expansions around $y=1-x$ 
\begin{eqnarray}
F(x) = \sum_{n=0}^\infty a(n) y^n
\end{eqnarray}
one obtains expansion coefficients $a(n)$, which serve as input sets for the method of 
arbitrary high moments, cf.~Ref.~\cite{Blumlein:2017dxp}. Using these sets of coefficients 
for the MIs one obtains the corresponding series expansion for the amplitudes, labeled by 
the color factors, zeta values \cite{Blumlein:2009cf}, and powers in the dimensional 
parameter $\ep$ from $O(1/\ep^3)$ to $O(\ep^0)$. 
Throughout the paper, the following constants appear\footnote{In the master integrals
multiple zeta values \cite{Blumlein:2009cf} up to {\sf w = 9} and $\zeta_{11}$ 
contribute, 
cf.~\cite{Blumlein:2019oas}, most of which cancel in the amplitude.}
\begin{equation}
  l_2 = \ln(2)\,, \,\,\,\,\,\, \zeta_k = \zeta(k) = \sum_{l=1}^\infty \frac{1}{l^k},~k \geq 2\,, 
\,\,\,\,\,\, 
a_n = \Li_n\left(\frac{1}{2}\right),~n \geq 4\,,  
\end{equation}
where $\zeta(k)$ is the Riemann $\zeta$ function evaluated at integer values 
$k \geq 2$, and
\begin{equation}
\Li_n(x) = \sum_{k=1}^{\infty} \frac{x^k}{k^n}\,, \,\,\,\,\,\, |x| \leq 1,
\end{equation}
are the classical polylogarithms, \cite{LEWIN1,LEWIN2,Devoto:1983tc}.

In the following we describe the basic tactic we used to compute these form factors.
Their amplitudes are given by 417 MIs,
\begin{equation}\label{Equ:LinComMi}
F_{I;j}(s)=c_1(\ep,s)M_1(\ep,s)+\dots+c_{417}(\ep,s)M_{417}(\ep,s)
\end{equation}
with rational functions $c_i(\ep,s)$ in $s$ and $\ep$.
A crucial property is that the form factors and the underlying master integrals can be 
given as a Laurent series expansion in $\ep$, starting from $O(\ep^{-3})$,
and as power series in $s$. 
Since the coefficients $c_i(\ep,s)$ also contain coefficients with factors $1/\ep^k$ 
with $-3\leq k \leq 10$,
one needs the $\ep$-expansion of the the MIs up to $\ep^k$ to determine the first four 
coefficients of the $\ep$-expansion of the form factors.
Besides the representation~\eqref{Equ:LinComMi}, 
the IBP reduction techniques delivers a coupled system 
of first-order linear differential 
equations
in terms of the master integrals. In particular, given a finite number of initial values 
(the first coefficients of the expansions in $s$ and $\ep$),
one can determine all the coefficients in $s$ up to a certain order in $\ep$.
Given this available representation~\eqref{Equ:LinComMi} together with the coupled system, 
the overall strategy of this article is based on the
large moment method~\cite{Blumlein:2017dxp}, which can be summarized as follows.

\begin{enumerate}
\item[I] Compute sufficiently many coefficients in $s$, say $\rho$, up to a certain order in $\ep$ of the series representations of the master integrals
  by using the coupled system together with the available initial values; see Section~\ref{Sec:LargeMoments}.

\item[II] Plug in the truncated power series solutions of the MIs into~\eqref{Equ:LinComMi} and expand the expressions to derive $\rho$ coefficients in $s$
  up to a certain order in $\ep$ of the series representation of the $F_I$'s. 
Note that the coefficients are given as a linear combination of the constants 
\begin{eqnarray}
	&& \kappa_1 = 1, \,\,\,\,\,\, \kappa_2 = \zeta_2, \,\,\,\,\, 
\kappa_3 = \zeta_3, \,\,\,\,\,\,\, \kappa_4 = l_2 \zeta_2, \,\,\,\,\,\, \kappa_5 = \zeta_2^2, \nonumber \\
	&& \kappa_6 = a_4, \,\,\,\, \kappa_7 = l_2^4, \,\,\,\,\, \kappa_8 
= l_2^2 \zeta_2, \,\,\,\, \kappa_9 = \zeta_2 \zeta_3, \,\,\,\, \kappa_{10} = \zeta_5,
	\label{eq:kappa-const}
\end{eqnarray}
which come from the initial values of the MIs and the color factors given in~\eqref{eq:color-factors}.
More precisely, one obtains a finite list of rational numbers for each form factor and each $C_i\kappa_j\ep^{k}$  with $1\leq i\leq 3$, $1\leq j\leq 10$ and $-3\leq k\leq 0$;
see Section~\ref{Sec:LargeMoments} for calculation details and Section~\ref{Sec:Guessing} for examples.

\item[III] For each finite list of rational numbers $(H(n))_{n=0}^{\rho-1}$ 
guess a linear recurrence and a linear differential equation, cf.~Refs.~\cite{SageOre,Blumlein:2009tj,GSAGE}
of the power series $h(s)=\sum_{n=0}^{\infty}H(n) s^n$; see Section~\ref{Sec:Guessing}.

\item[IV] Using this information utilize the recurrence solver of 
  \texttt{Sigma}~\cite{SIG1,SIG2,Schneider:2013zna} 
  in the setting of difference 
 rings~\cite{KARR,CS1,CS2,CS3,CS4,CS5,CS6,CS7,CS8,BRONSTEIN,ABRAMOV,
  CS9,CS10,CS11,CS12} to determine for which physical components the coefficients are expressible 
  in terms of iterative sums. Then transform the underlying functions of their power series 
  representation in terms of iterative integrals using the 
package \texttt{HarmonicSums}~\cite{
  Ablinger:2010kw,Ablinger:2013hcp}. Furthermore, 
techniques are necessary to handle finite binomial 
sums, cf.~Refs.~\cite{Ablinger:2014bra,Ablinger:2015tua}, cf.~Section~\ref{Sec:Guessing}. All those closed form representations can be straightforwardly 
  evaluated at any 
  expansion point using techniques described 
in~Refs.~\cite{Gehrmann:2001pz,Ablinger:2018sat,CPOLYF,Vollinga:2004sn}.

\item[V]
  For all contributions where the recurrences and differential equations are not first-order factorizable
  (i.e., the full solution space cannot be given in terms of iterative sums/integrals),
  perform analytic continuations and assemble the final result to derive representations of all the form factors at other expansion points; see Section~\ref{Sec:AnalyticCont}.
\end{enumerate}

We note that the required number $\rho$ of coefficients is determined by the guessing method.
If one fails to guess a recurrence or differential equation from the given number of 
coefficients in Step III,
one has to produce more coefficients in Steps~I and~II until one succeeds. 

Several ingredients of the calculation were already employed in the quarkonic computation of Ref.~\cite{Blumlein:2023uuq}.
The gluonic case is, however, considerably more involved. Because of this, a number of new 
techniques had to be developed in order to complete the calculation.
\section{Computing coefficients from the coupled systems}
\label{Sec:LargeMoments}

\vspace*{1mm}
\noindent
For the gluonic case of the form factor, we obtain a coupled system of 419 MIs that can be 
hierarchically structured into 147
recursively defined sub-systems obtained by  the integration-by parts reduction 
\cite{IBP1,IBP2,IBP3,IBP4,Chetyrkin:1981qh,Laporta:2001dd,CRUSHER}. The 
smallest sub-systems have dimension 1 
and the largest ones have dimension 10. In this way,
we can apply the large moment method in a bottom-up fashion for these reasonable sized 147 
sub-systems.
In addition, the physical expressions (scalar, pseudoscalar, vector, axial-vector)
in the different color factors are given by a linear combination of these MIs, 
as shown in Eq.~(\ref{eq:F-color-decomp}). 
After compactification of terms~\cite{RefinedLM2:24}, the input obtained amounts to about 5~GB.

In the following we focus on such a given coupled system of first-order linear differential equations in terms of
the unknown MIs $M_1(s),\dots,M_{\lambda}(s)$ with $1\leq \lambda\leq 10$ that have a power series representation
$M_i(s)=\sum_{n=0}^{\infty} G_i(n)s^n$. Then
the large moment method introduced in~\cite{Blumlein:2017dxp} deals with the problem to compute efficiently the coefficients $G_i(n)$ with $0\leq n<\rho$
for some large $\rho$; in simpler cases $\rho$ may range between 500 and 10000. 
We note that the inhomogeneous 
part depends on the master integrals coming from the sub-sectors below. Here 
we assume that we have computed already the first $\rho$ coefficients of their 
power series expansions, either by other simpler methods or our large moment 
method 
applied recursively.
The underlying steps can be summarized as follows. 
\begin{enumerate}
  \item[I.1] We uncouple the given system. Among the available algorithms, we 
    mainly use
    Gau\ss{}' and Z\"urcher's algorithm
    implemented in the \texttt{OreSys} package~\cite{ORESYS}. For further 
    details we refer 
    to Refs.~\cite{Zurcher1994,BCP13}.
	
  \item[I.2] Based on the different uncoupling algorithms,
    we obtain one (or several) scalar differential equations 
    \begin{equation}\label{Equ:DELM}
	b_0(s)M(s)+b_1(s)\frac{d}{ds}M(s)+\dots+b_{\lambda}(s)\frac{d^{\lambda}}{ds^{\lambda}}M(s)=r(s)
    \end{equation}
    with $M(s)\in\{M_1(s),\dots,M_{\lambda}(s)\}$, polynomials $b_i(s)$ in $s$ and a given right hand side $r(s)$ that is given by a linear combination in terms of the
    already handled master integrals. 
    Plugging $M(s)=\sum_{n=0}^{\infty} G(n)s^n$ into this equation and performing coefficient comparison with respect to $s^n$ leads to a linear recurrence relation
    \begin{equation}\label{Equ:RECLM}
	a_0(n)G(n)+\dots+a_{\delta}(n)G(n+\delta)=R(n)
    \end{equation}
    of some order $\delta$ with polynomials $a_i(n)$ in $n$ where the coefficients $R(n)$ for $n<\rho$ are given explicitly.
	
  \item[I.3] Without loss of generality, we may assume that $a_{\delta}(n)\neq0$ for all $n\geq\delta$;
    otherwise increase $\delta$ such that this is the case.
    Suppose in addition that we are given $\delta$ initial values for $G(n)$. Then we use the recurrence~\eqref{Equ:RECLM} to prolong the sequence $G(n)$ to $n<\rho$
    in linear time with respect to the number of operations in\footnote{We note that the size of the rational numbers grows significantly for large $n$, meaning arithmetic operations
    require non-constant time in terms of processor words. We omit a formal complexity analysis and instead provide later computation times to illustrate the computational
    challenge.} $\mathbb Q$. 
	
  \item[I.4] Any of the available uncoupling algorithms from Step~I.1 allows one to express the remaining unknown MIs in terms of a linear combination
    of the determined MIs (including their differentiated versions) from the previous step (and the already solved MIs from previous recursions).
    Thus expanding those expressions leads to a power series representation where the first $\rho$ coefficients are given explicitly.
\end{enumerate}

An extra complication is that the coupled system contains also the 
dimensional parameter $\ep$ and one is interested in a Laurent series expansion of the $M_i(\ep,s)$
not only in $s$ but but also in $\ep$. Here two 
refined methods have been elaborated 
in~Refs.~\cite{Blumlein:2019oas,RefinedLM1:20}
where in both cases one ends up at a linear recurrence of the form~\eqref{Equ:RECLM}. 
Depending on the choice of the strategy, the coefficients
are free of $\ep$, i.e., we simply follow Step~I.3 above. If they depend on 
$\ep$ one may use the tools described in~Ref.~\cite{BKSS:12} to prolong the sequence
for the different $\ep$-contributions.

To fine-tune the large moment method, we carried out the different systems systematically 
by applying all the available uncoupling methods, in our case Gau\ss{} and Z\"urcher,
and the available strategies to handle the $\ep$-expansions in order to select the best 
strategy such that the order $\delta$ of the derived recurrences and
the required $\ep$-order of master integrals is minimized simultaneously; 
for a careful study we refer also to~\cite{Fadeev:25}.

After applying the four steps I.1--I.4 from above iteratively to the 147 recursively defined 
sub-systems, one deals with the physical expression coming from
the IBP reductions as carried out in Step~II: one plugs in the explicitly given expansions of the master integrals up to the order $\rho$ into
the physical expression and expands it to get the first $\rho$ coefficients of the $s$-expansion. Examples of the explicit expansions of the
form factors $F_I$ coming from Step~II will be given in Section~\ref{Sec:Guessing}.
\subsection{From few to many coefficients}\label{Sec:InitialValuesVersusRHS}

\vspace*{1mm}
\noindent
In order to prolong the coefficients to the desired number $\rho$ using the recurrence~\eqref{Equ:RECLM} in Step~I.2, at least $\delta$ initial values have to be provided.
If $\delta$ is too large\footnote{Whereas about $20$ initial values can easily be computed for simpler integrals, the task already becomes demanding at $10$ values for more involved cases.}, their direct calculation becomes computationally prohibitive.
Thus keeping $\delta$ as small as possible is crucial when deriving the recurrence~\eqref{Equ:RECLM} from the differential equation~\eqref{Equ:DELM}.

\textbf{Strategy 1: Minimizing the order}.  We leverage the fact that the order $\delta$ is bounded by (and generically equals)
$$\delta\leq\max_i\deg_s(b_i)+\lambda;$$
compare~Refs.~\cite{KP:11,Kaue2023}.
To minimize $\delta$, a straightforward approach is to reduce the degrees of the coefficients $b_i(s)$ by canceling their greatest common divisor.
Further refinements from Refs.~\cite{Blumlein:2019oas,RefinedLM1:20} reduced these 
potential orders by roughly 40, bringing the required number of initial values
into a manageable computational range.

We note that minimizing the order $\delta$ is made possible by shifting the extra factors from the left-hand side of~\eqref{Equ:DELM} to the denominators of the right-hand side,
$r(s)$. However, this introduces more complicated rational functions multiplying the already-evaluated master integrals. Consequently, the power series expansion of $r(s)$
required to obtain $R(n)$ for $n=0,\dots,\rho-1$ becomes significantly more involved. 
We note that these complications could be managed for expansions up to $\rho=12000$
using the techniques described in~Ref.~\cite{RefinedLM2:24}.

For the different contributions the number $\rho$ was chosen big enough to succeed in guessing 
afterwards a linear recurrence/differential equation of the physical problem coming from the 
coefficients when performing Step~II introduced above. Recursions of different complexity
need input sets of different size, ranging from $\rho=500$ to $\rho=12000$ for all but the 
three cases without $\zeta$-weights, which required about 6 months of parallel running 
using 10--12 processes.

However, we needed much more coefficients to handle the remaining parts. In a 
first attempt we calculated $\rho=25000$ coefficients.
This challenge forced us to invert the above strategy entirely.

\textbf{Strategy 2: Simplifying the right-hand side}. We clear all denominators in $r(s)$ by multiplying both sides by a large polynomial, which increased in the worst cases
the degrees of the $b_i(s)$ up to 200. This yields a linear recurrence~\eqref{Equ:RECLM} with $\lambda\sim200$. On the other side, $r(s)$ is now given as a linear combination
of master integrals with simple polynomial coefficients in $s$ (or $s,\varepsilon$). Consequently, expanding $r(s)$ to obtain $R(n)$ for $n=0,\dots,\rho-1$ becomes significantly
more efficient. Although the prolongation requires $\lambda\sim200$ initial values and takes longer due to the larger recurrence, its complexity remains linear in the number of field operations.
For instance, by taking $\rho=25000$ this prolongation was completed in just a few 
hours.

\textbf{Combining Strategy~1 and~2:} The true strength of our method lies in combining 
these two approaches. First, we compute $\rho=500$ values for the full problem within a 
day using the first approach, leveraging the initial values available from direct calculations.
Second, we utilize these 500 coefficients as  the initial values for
the second approach to scale the computation up to $\rho=25000$.

We remark that Strategy~2 can be also applied to speed up  Step~II: clearing denominators 
of the physical expression with a common huge factor $d(s,\ep)$ enables one to compute the 
expansion in $s$ and $\ep$ by simple polynomial arithmetic. Only in the end, one divides 
the found truncated polynomial by $d(s,\ep)$ and computes the expansion
in $s$ and $\ep$ following the ideas presented in~Ref.~\cite{RefinedLM2:24}.
To compute $\rho=25000$ coefficients of all the MIs took about 11 months.

In our original implementation of the package \texttt{SolveCoupledSystem} 
\cite{Ablinger:2016wlr,Blumlein:2019oas,Blumlein:2019hfc}
implemented for {\tt Mathematica}, standard parallelization has been treated that  
assumes that
the appearing MIs can be stored directly in memory, a strategy that successfully sustained all our previous calculations. However, scaling up to 147 systems
introduced non-trivial memory management challenges, as these systems comprise 419 MIs yielding 25,000 total values that require 5.6 TB of storage.
The final values are particularly demanding, with each requiring approximately 
130~kB of memory to store their 150,000 to 157,000 digits,
while the right hand side of the latest systems depend on 200 to 300 MIs simultaneously. 
To handle 
the 20 most complicated systems,
new computational techniques became necessary. Parallelization was distributed across 
multiple computers and reorganized for any given right hand side to restrict memory usage
to a maximum of up to 10 MIs per {\tt Mathematica} call. Furthermore, intermediate 
sub-results, 
which are significantly larger than the final right hand side,
were written directly to hard disks, demanding meticulous coordination of disk space and subsequent manual combination. Finally, employing different uncoupling strategies,
such as Gau\ss{} and Z\"urcher, introduced varying communication requirements that had to 
be partially managed through manual intervention.
\subsection{Prolonging to larger coefficients}

\vspace*{1mm}
\noindent
We applied the techniques described in Section~\ref{Sec:InitialValuesVersusRHS} to compute 
$\rho=25000$  coefficients for the constant contributions.
This number proved to be sufficient to guess the required recurrence relations for all
but the color factor $C_F^3$ for $F_{a,1}(s)$, see 
Eqs.~\eqref{Agidecomp1} and \eqref{eq:F-color-decomp}. 
Rather than restarting the entire pipeline, we leveraged the existing coefficients to 
efficiently extend the computation to $\rho=40000$  using the strategy outlined 
below, which allowed us to guess the
yet missing differential equations.

\begin{enumerate}    
\item[(i)] Using the 24000 coefficients, we guessed as many recurrence relations as 
 possible for the 417 master integrals in 2 months of real time.
 For the remaining systems, only the last two $\varepsilon$-orders for 2 systems, 
 and the final $\varepsilon$-order for 10 systems, could not be guessed.
	
\item[(ii)] With the successfully guessed recurrences, we generated the 
coefficients up to $\rho=40000$ in about 7 days on several machines. 
	
	\item[(iii)] For the remaining 12 systems, we computed the values up to 
$\rho=40000$ exclusively for the missing $\varepsilon$-contributions, while 
leveraging the lower $\varepsilon$-contributions obtained from the previous step. This 
particular calculation could not be performed in a fully automated fashion;
          it required manual intervention, custom job distribution of sub-problem 
across different machines, and meticulous memory management. 
In particular,
          intermediate results had to be strategically distributed and 
stored across multiple hard drives. This step amounted to
1.3 months of real run time.
\end{enumerate}

In its final form, the complete dataset required approximately 5.5 TB of storage in 
compressed form. Combining the master integrals to extract the 40000 coefficients
of $C_F^3$ for $F_{a,1}(s)$ took 3 weeks using parallelization. Naturally,
managing these memory demands and parallelization workflows required significant 
additional effort compared to our previous 24000 coefficient calculations, as well as an
increased level of manual intervention. The complete computation of the 40000 
coefficients 
for the missing axial-vector contribution spanned 3.5 months running in parallel. 

Ultimately, it turned out that only 26000 coefficients were strictly necessary to 
successfully guess the required recurrence relation.

\section{Recurrences and differential equations from the low energy expansions 
of the form factors}\label{Sec:Guessing}

\vspace*{1mm}
\noindent
The initial values for all differential equations considered in this project are
given at $s = 0$, resp. $x = 1$, the low energy limit of the form factors. 

Once we have computed the expansion of the master integrals at this point in Step~I, we 
can combine them in Step~II to obtain
the low energy expansion of the unrenormalized form factors. In particular, we can obtain 
the expansions about $s=0$ of the
form factors on the right-hand side of Eqs. (\ref{Vgidecomp1}--\ref{Agidecomp1}) and Eqs. (\ref{Sdecomp1}--\ref{Pdecomp1}).
These expansions up to ${O}(\ep^0)$ are of the form
\begin{equation}
  F_I^{s \rightarrow 0}(s) = \sum_{i=-3}^0 \ep^i \sum_{j=1}^3 C_j \sum_{k=0}^\infty 
c_{I;i,j,k}^{(0)} s^k,
\label{eq:expansion-s=0}
\end{equation}
where the color factors $C_j$ are given in Eq.~(\ref{eq:color-factors}), while the coefficients $c_{I;i,j,k}^{(0)}$ are rational linear combinations of the
following list of constants\footnote{The label ``(0)'' in $c_{I;i,j,k}^{(0)}$ is used to emphasize that the expansion is around $s=0$. We will later
obtain expansions around other values of $s$, and the label will be changed accordingly.},
\begin{equation}
  \left\{1, \, \zeta_2, \,  \zeta_3, \, l_2 \zeta_2, \, \zeta_2^2, 
\, a_4, \, l_2^4, \, l_2^2 \zeta_2, \, \zeta_2 \zeta_3, \, \zeta_5 \right\}.
  \label{eq:MZV-list}
\end{equation}
Using a finite but large number $\rho$ of rational numbers, 
we executed an automated 
guessing strategy to find the underlying recurrence relations
and linear differential equations. 

In our specific setup, it became evident that 
significantly fewer coefficients were required to guess the recurrence relations
than the corresponding differential equations. Consequently, we adopted a staged approach: 
we first guessed the recurrences, used them to generate additional coefficients, like  
extending 24000 initial coefficients needed for the recurrence to 40000 values needed for 
the differential equation. The axial-vector case stood out as an exception.

The guessing was performed using the \texttt{ore\_algebra} package 
in {\tt Sage}, cf.~Refs.~~\cite{SageOre,GSAGE}. For the most demanding case requiring 
around $25300$ values to guess the linear recurrence and 37700 values to guess the linear 
differential equation we utilized a parallel version of the \texttt{ore\_algebra} 
package, which was used at the supercomputer {\tt MACH-2} at JKU to exploit 1728 processor 
cores 
and 20 TB of global shared memory. In one of the most computationally involved 
differential equation runs, 
the guessing algorithm utilized 24,420 coefficients across 
800 cores, with each core computing one homomorphic image, requiring 30 hours via 
parallelization. Subsequently, a further optimized version of the \texttt{ore\_algebra} 
package, allowed us to rerun this hardest case in approximately one day on a single fast 
workstation at RISC, parallelizing twelve times.\footnote{In the present 
installation we use the {\tt Sage} linear solver {\tt linbox\_noefd}.}

If we rename the constants (\ref{eq:MZV-list}) as given in~\eqref{eq:kappa-const} 
the coefficients $c_{I;i,j,k}^{(0)}$ can be written as
\begin{equation}
  c_{I;i,j,k}^{(0)} = \sum_{l=1}^{10} r_{I;i,j,l,k}^{(0)} \kappa_l,
\end{equation}
where the $r_{I;,i,j,l,k}^{(0)}$'s are rational numbers. Introducing this equation in (\ref{eq:expansion-s=0}), 
we can rewrite the form factors as
\begin{equation}
  F_I^{s \rightarrow 0}(s) = \sum_{i=-3}^0 \ep^i \sum_{j=1}^3 C_j \sum_{l=1}^{10} \kappa_l f_{I;i,j,l}^{(0)}(s),
\label{eq:F2fjl}
\end{equation}
where the expansions around $s=0$ of the functions $f_{I;i,j,l}(s)$ are given by
\begin{equation}
  f_{I;i,j,l}^{(0)}(s) = \sum_{k=0}^\infty r_{I;i,j,l,k}^{(0)} s^k.
\label{eq:fjl}
\end{equation}

The differential equations obeyed by these functions and obtained through guessing algorithms will be of the form
\begin{equation}
  p_0(s) f_{I;i,j,l}^{(0)}(s) + p_1(s) \frac{d}{ds} f_{I;i,j,l}^{(0)}(s) + p_2(s) \frac{d^2}{ds^2} f_{I;i,j,l}^{(0)}(s) + \cdots + p_r(s) \frac{d^r}{ds^r} f_{I;i,j,l}^{(0)}(s) = 0,
  \label{eq:diffeq-per-const}
\end{equation}
where $p_0(s), \, p_1(s), \ldots, p_r(s)$ are polynomials in $s$.
We will therefore have a recurrence relation and
a differential equation for each one of the constants in (\ref{eq:MZV-list}) appearing in each color factor of each form factor.
In this way, we have moved from the problem of solving differential equations obeyed by the master integrals,
as it was done in Refs.~~\cite{Fael:2022miw,Egner:2022jot,Fael:2022rgm,Fael:2023zqr} and 
many other perturbative calculations,
to the problem of solving differential equations derived from expansion coefficients.
This is the same approach we used in Ref.\cite{Blumlein:2023uuq} for the quarkonic case, 
now applied to the more involved gluonic contributions.
As we will see in the next sections, this method allows us to get to such a high
precision to determine the form factors in the high energy limit analytically.
In earlier calculations this problem was solved purely numerically at a sufficient 
level of precision
in~Refs.~\cite{Fael:2022miw,Egner:2022jot,Fael:2022rgm,Fael:2023zqr}.

In order to guess closed-form differential equations, it is sometimes necessary 
to drop the first few rational numbers, which leads to inhomogeneites.
This inhomogeneity can subsequently be eliminated, yielding a differential equation of 
the form (\ref{eq:diffeq-per-const}) at the expense of increasing the order of the 
differential equation by one.
\begin{table}[h]
\begin{center}
\begin{tabular}{|c|c|c|c|c|c|c|c|c|c|}
\hline    
          &             & $\zeta_2^2$ &   $l_2^2 \zeta_2$   &  $l_2^4$   &  $a_4$  &  $l_2 \zeta_2$ & $\zeta_3$ & $\zeta_2$ &   1 \\ 
\hline
          & $C_F^3$     &     14      &          14        &     14     &    14    &       33      &     57    &     59   &  96  \\
$F_s$     & $C_A C_F^2$ &     14      &          14        &     14     &    14    &       33      &     56    &     57   &  93  \\ 
          & $C_A^2 C_F$ &     13      &          13        &     11     &    11    &       32      &     54    &     55   &  90  \\
\hline
          & $C_F^3$     &     13      &          13        &     12     &    12    &      33      &     56    &     58   &  96  \\
$F_p$     & $C_A C_F^2$ &     14      &          14        &     13     &    13    &      33      &     54    &     56   &  92  \\ 
          & $C_A^2 C_F$ &     11      &          11        &     9      &     9    &      30      &     49    &     51   &  85  \\
\hline
          & $C_F^3$     &     14      &          14        &     13     &    13    &      36      &     58    &     60   &  101 \\
$F_{v,1}$  & $C_A C_F^2$ &     14      &          14        &     13     &    13    &      36      &     58    &     61   &  100 \\ 
          & $C_A^2 C_F$ &     14      &          14        &     12     &    12    &      35      &     58    &     58   &  95  \\
\hline
          & $C_F^3$     &     14      &          14        &     13     &    13    &      34      &     57    &     59   &  98  \\
$F_{v,2}$  & $C_A C_F^2$ &     14      &          14        &     13     &    13    &      34      &     57    &     59   &  97  \\ 
          & $C_A^2 C_F$ &     14      &          14        &     12     &    12    &      34      &     55    &     56   &  92  \\
\hline
          & $C_F^3$     &     14      &          14        &     14     &    14    &      36      &     58    &     62   &  102 \\
$F_{a,1}$  & $C_A C_F^2$ &     14      &          14        &     14     &    14    &      36      &     58    &     62   &  101 \\ 
          & $C_A^2 C_F$ &     14      &          14        &     12     &    12    &      35      &     58    &     59   &   96 \\
\hline
          & $C_F^3$     &     13      &          13        &     12     &    12    &      33      &     56    &     58   &  96  \\
$F_{a,2}$  & $C_A C_F^2$ &     14      &          14        &     13     &    13    &      33      &     54    &     56   &  92  \\ 
          & $C_A^2 C_F$ &     11      &          11        &     9      &     9    &      30      &     49    &     51   &  85  \\
\hline
\end{tabular}
\caption{\sf \small Orders of the differential equations obtained from the rational coefficients for the different color factors and different constants
  in (\ref{eq:MZV-list}) for each form factor (after removing inhomogeneities, if needed). In the case of the weight {\sf w}~=~5 constants, $\zeta_2 \zeta_3$ and $\zeta_5$,
  the orders are always equal to 2.}
\label{orderstable}
\end{center}
\end{table}

The recursion relations can now be used to generate a much larger amount of rational 
numbers in Eq.~(\ref{eq:fjl}). We derived 100000 such rational numbers for all of the 
constants in (\ref{eq:MZV-list}), for each form factor and color factor. 
This allowed us to evaluate the functions $f_{I;i,j,l}^{(0)}(s)$ with extremely high 
precision within their radius of convergence, which is important for the next steps.

All pole 
terms obey first order factorizing difference equations.
Using the recurrence solver in the setting of difference 
rings~\cite{KARR,CS1,CS2,CS3,CS4,CS5,CS6,CS7,CS8,BRONSTEIN,ABRAMOV,CS9,CS10,CS11,CS12}
of the summation package \texttt{Sigma}~\cite{SIG1,SIG2}, all coefficients can be 
expressed in terms of cyclotomic harmonic sums~\cite{Ablinger:2011te}.
Using the package~\texttt{HarmonicSums}~\cite{Ablinger:2010kw,Ablinger:2013hcp} their power series representations can be rewritten
in terms of harmonic polylogarithms in $x$, \cite{Remiddi:1999ew}, or in Mellin space in
harmonic sums \cite{Vermaseren:1998uu,Blumlein:1998if}.

At ${O}(\ep^0)$, the differential equations are first-order factorizable 
in the case of the constants $\zeta_5$ and $\zeta_2 \zeta_3$.
For the remaining constants in (\ref{eq:MZV-list}), the differential equations are 
not first-order factorizable.

The degree of the polynomials and the order $r$ of the differential equations 
(\ref{eq:diffeq-per-const}) will in general depend on the values of $I$, $i$, $j$ and $l$.
In Table~\ref{orderstable}, the orders $r$ of the differential equations  
obtained for the constants in (\ref{eq:MZV-list}) for each color factor and each form 
factor are given, after inhomogeneities were removed.
In Table~\ref{degreestable}  we show the corresponding highest degrees of the 
polynomials.

\begin{table}[!h]
\begin{center}
\begin{tabular}{|c|c|c|c|c|c|c|c|c|c|}
\hline    
          &             & $\zeta_2^2$ & $l_2^2 \zeta_2$ & $l_2^4$ &  $a_4$  & $l_2 \zeta_2$ & $\zeta_3$ & $\zeta_2$ &   1 \\ 
\hline
          & $C_F^3$     &    202      &      202       &   200   &   200   &     1045      &    1447   &   1434   &  3708  \\
$F_s$     & $C_A C_F^2$ &    204      &      204        &   200  &   200   &     1047      &   1420    &   1379   &  3544  \\ 
          & $C_A^2 C_F$ &    187      &      187        &  147   &   147   &      989      &   1324    &   1297   &  3248  \\
\hline
          & $C_F^3$     &    179      &      179        &  166   &   166   &     1057      &   1429    &   1428   &  3712  \\
$F_p$     & $C_A C_F^2$ &    198      &      198        &  183   &   183   &     1065      &   1353    &   1316   &  3419  \\ 
          & $C_A^2 C_F$ &    139      &      139        &   95   &    95   &      875      &   1081    &   1083   &  2756  \\
\hline
          & $C_F^3$     &    216      &      216        &  201   &   201   &     1210      &   1537    &   1531   &  4033 \\
$F_{v,1}$  & $C_A C_F^2$ &    218      &      218        &  201   &   201   &     1212      &   1553    &   1547   &  3978 \\ 
          & $C_A^2 C_F$ &    212      &      212        &  160   &   160   &     1127      &   1468    &   1407   &  3613 \\
\hline
          & $C_F^3$     &    194      &      194        &  181   &   181   &     1072      &   1448    &   1450   &  3814  \\
$F_{v,2}$  & $C_A C_F^2$ &    196      &      196        &  181   &   181   &     1074      &   1447    &   1448   &  3762  \\ 
          & $C_A^2 C_F$ &    190      &      190        &  152   &   152   &     1052      &   1367    &   1338   &  3428  \\
\hline
          & $C_F^3$     &    216      &      216        &  214   &   214   &     1208      &   1549    &   1570   &  4078 \\
$F_{a,1}$  & $C_A C_F^2$ &    218      &      218        &  214   &   214   &     1210      &   1552    &   1568   &  4026 \\ 
          & $C_A^2 C_F$ &    212      &      212        &  172   &   172   &     1127      &   1472    &   1440   &  3684 \\
\hline
          & $C_F^3$     &    179      &      179        &  166   &   166   &     1057      &   1429    &   1428   &  3712 \\
$F_{a,2}$  & $C_A C_F^2$ &    198      &      198        &  183   &   183   &     1065      &   1353    &   1316   &  3419 \\ 
          & $C_A^2 C_F$ &    139      &      139        &   95   &    95   &     1081      &   1081    &   1083   &  2756 \\
\hline
\end{tabular}
\caption{\sf \small Highest degrees of the polynomials appearing in the differential equations obtained from the rational coefficients
  for the different color factors and different constants in (\ref{eq:MZV-list}) for each form factor (after removing inhomogeneities, if needed).
  In the case of the constants $\zeta_2 \zeta_3$ and $\zeta_5$, these degrees are all equal to 6.}
\label{degreestable}
\end{center}
\end{table}

The singular points of the differential equations also depend on the choice of constant in (\ref{eq:MZV-list}).
The differential equations associated to the weight {\sf w}~=~4 constants have real valued singularties at $s=0$, $s=-1$, $s=-4$, $s=3$ and $s=4$. In the case of $l_2 \zeta_2$,
in addition to the previous ones, we have also singularities at $s=-1/2$ and $s=1$, while 
in the case of $\zeta_3$, we must add a singularity at $s=16$, and
in the cases of $\zeta_2$ and $\kappa_1 = 1$, also a singularity at $s=16/3$ pops up.
In total, there are five singularities in the case of $\zeta_2^2$, $l_2^2 \zeta_2$, $l_2^4$ and $a_4$, seven singularities in the case of $l_2 \zeta_2$,
eight singularities for $\zeta_3$, and nine singularities in the cases of $\zeta_2$ and $\kappa_1 = 1$.
Of all these singularities, only the ones at $s=4$ and $s=16$ are physical, 
corresponding to the two-particle threshold and four-particle pseudo-threshold, respectively.\footnote{All the contributing 
differential equations have as well a singularity as $s \rightarrow \infty$.} 
The remaining singularities are spurious, and arise only as a consequence of having split the form factors in terms of
the functions $f_{I;i,j,l}^{(0)}(s)$ associated to the constants in (\ref{eq:MZV-list}).
After we combine everything in (\ref{eq:F2fjl}), these spurious singularities disappear.

The simplest expansions for the form factors are obtained in the limit $s \rightarrow 0$.
According to Eqs.~(\ref{eq:F2fjl}) and (\ref{eq:fjl}), the ${O}(\ep^0)$ parts of the unrenormalized form factors are given by 
\begin{equation}
\label{eq:x1}
F_I^{s \rightarrow 0}(s) = \sum_{j=1}^3 C_j \sum_{l=1}^{10} \kappa_l f^{(0)}_{I;0,j,l}(s) 
=
\sum_{j=1}^3 C_j \sum_{l=1}^{10} \sum_{k=0}^\infty r^{(0)}_{I;0,j,l,k} \kappa_l s^k,
\end{equation}
where the $C_j$'s are the color factors given in Eq.~(\ref{eq:color-factors}), while the 
coefficients $r^{(0)}_{I;0,j,l,k}$ are rational numbers.

Before we show an explicit example of such an expansion, 
we should talk about the solvable parts of the form factors.
For two of the constants in Eq.~(\ref{eq:kappa-const}), namely, $\kappa_9 = \zeta_2 
\zeta_3$ and $\kappa_{10} =  \zeta_5$,
the corresponding differential equations in the variable $x$ for the functions 
$f_{I;0,j,l}(s)$ in Eq.~(\ref{eq:F2fjl}) were solved analytically.
As might have been expected, due to the high transcendentality of these constants, the 
solutions turned out to be very simple, depending only on $\ln(x)$ and rational functions 
in $x$.
In the vector case, setting $N_c=3$, for the corresponding contributions to $F_{v,1}(s)$ 
at 
${O}(\ep^0)$ we obtained 

\begin{eqnarray}
  \sum_{j=1}^3 C_j \sum_{l=9}^{10} \kappa_l f_{v,1;0,j,l}(s) &=&
64 \Biggl\{-\zeta_2 \zeta_3 \left(
                \frac{2018-7743 x-3634 x^2-7743 x^3+2018 x^4}{27 (1-x) x (1+x)} \ln(x)
\right. \nonumber\\ && \left.
                +\frac{1703-8712 x+1703 x^2}{9 x}
        \right) 
        + \zeta_5 \left(\frac{4313-27032 x+4313 x^2}{108 x}
\right. \nonumber\\ && \left.
        -\frac{5 \left(502-2041 x+2196 x^2-2041 x^3+502 x^4\right)}
{54 (1-x) x (1+x)} \ln(x) \right)\Biggr\},
\label{eq:solvable-fv1}
\end{eqnarray}
while the corresponding contributions to $F_{v,2}(s)$ are given by\footnote{The 
corresponding 
expressions for the other form factors are given in an ancillary file.} 
\begin{eqnarray}
  \sum_{j=1}^3 C_j \sum_{l=9}^{10} \kappa_l f_{v,2;0,j,l}(s) &=&
   64     \Biggl\{-\zeta_2 \zeta_3 \left(
          \frac{2018+3940 x+5773 x^2+3940 x^3+2018 x^4}{54 (1-x) x (1+x)} \ln(x)
\right. \nonumber\\ && \left.   
          +\frac{26561+49264 x+26561 x^2}{216 x}
        \right)
        +\zeta_5 \left(\frac{13801+25072 x+13801 x^2}{432 x}
\right. \nonumber\\ && \left.               
          -\frac{5 \left(502+1028 x+799 x^2+1028 x^3+502 x^4\right)}{108 (1-x) 
x (1+x)} \ln(x)
       \right)\Biggr\}.
\label{eq:solvable-fv2}
\end{eqnarray}

The poles in $\ep$ in Eq.~(\ref{eq:F2fjl}) can also be solved exactly as functions of $x$. However, the resulting expressions are generally considerably
more involved than those in Eqs.~(\ref{eq:solvable-fv1}--\ref{eq:solvable-fv2}), with their complexity increasing with the order in $\ep$.
The ${O}(\ep^{-3})$ terms have a structure similar to that of the results above, being expressed solely in terms of $\ln(x)$ and rational functions of $x$.
At ${O}(\ep^{-2})$, harmonic polylogarithms of weight up to {\sf w}~=~3 appear, 
whereas the ${O}(\ep^{-1})$ terms involve harmonic polylogarithms of weight up to {\sf w}~=~5. 

The exact solutions in $x$ for the $\zeta_2\zeta_3$ and $\zeta_5$ contributions at ${O}(\ep^0)$, such as those given in Eqs.~(\ref{eq:solvable-fv1})--(\ref{eq:solvable-fv2}),
together with the exact solutions for the poles in $\ep$, are valid throughout the entire kinematic range\footnote{For this reason we removed the labels ``(0)'' or ``($s_0$)''
from $f_{I;0,j,l}(s)$ in Eqs.~(\ref{eq:solvable-fv1})--(\ref{eq:solvable-fv2})}.
They can therefore be expanded about any value of $s$, including $s=0$,
by using Eq.~(\ref{eq:x2s}).

When reconstructing the form factors $F_{V,1}(s)$ and $F_{V,2}(s)$ from Eq.~(\ref{Vgidecomp1}), or $F_{A,1}(s)$ and $F_{A,2}(s)$ from Eq.~(\ref{Agidecomp1}),
it is important to take into account that the projector coefficients $g_{I,i}^{(k)}$ in Eqs.~(\ref{eq:gV1}--\ref{eq:gA2}) depend on $\ep$.
Consequently, after expanding these coefficients in $\ep$, the contributions of order ${O}(\ep^i)$ with $i \leq 0$ contribute to the finite, ${O}(\ep^0)$, terms.

As an illustrative example, we present the expansion of the vector current form factor 
$F_{V,1}(s)$ up to ${O}(s^5)$.
It is obtained by combining the expansions of the non-solvable contributions in 
Eq.~(\ref{eq:F2fjl}), corresponding to the constants $\kappa_l$ with $1 \leq i \leq 8$ 
in Eqs.~(\ref{eq:kappa-const}), with the expansions of the exact $\zeta_2\zeta_3$ and $\zeta_5$ 
solutions, together with the contributions originating from the poles in $\ep$.
Setting again $N_c=3$, we obtain
\begin{eqnarray}
\label{eq:xx1}
\lefteqn{F_{V,1}^{s \rightarrow 0} =} \nonumber\\ &&
64 \Biggl\{
\frac{274321}{324}
+\frac{589477}{2592} \zeta_2
+\frac{3679}{27} l_2 \zeta_2
-\frac{619}{18} l_2^2 \zeta_2
+\frac{28873}{1440} \zeta_2^2
-\frac{3079}{108} \zeta_3
-\frac{349}{9} \zeta_2 \zeta_3
+\frac{713}{27} \zeta_5
\nonumber\\ && 
-\frac{2099}{27} a_4
-\frac{2099}{648} l_2^4
+s \Biggl[
        -\frac{30176683}{139968}
        +\frac{658}{27} a_4
        +\frac{329}{324} l_2^4
        -\frac{1381189}{139968} \zeta_2
\nonumber\\ && 
        -\frac{31435}{648} l_2 \zeta_2
        +\frac{350}{27} l_2^2 \zeta_2
        +\frac{41711}{2430} \zeta_2^2
        +\frac{258683}{11664} \zeta_3
        -\frac{905}{108} \zeta_2 \zeta_3
        -\frac{1165}{432} \zeta_5
\Biggr]
\nonumber\\ && 
+s^2 \Biggl[
        -\frac{188800077257}{2799360000}
        +\frac{104717}{12150} a_4
        +\frac{104717}{291600} l_2^4
        -\frac{17227423687}{4898880000} \zeta_2
\nonumber\\ && 
        -\frac{181845473}{13608000} l_2 \zeta_2
        +\frac{315487}{85050} l_2^2 \zeta_2
        -\frac{43819}{272160} \zeta_2^2
        +\frac{935160107}{124416000} \zeta_3
        -\frac{7507}{6480} \zeta_2 \zeta_3
        -\frac{53}{144} \zeta_5
\Biggr]
\nonumber\\ && 
+s^3 \Biggl[
        -\frac{75271254677139169}{3629482214400000}
        +\frac{1053107}{297675} a_4
        +\frac{1053107}{7144200} l_2^4
        -\frac{566793123524251}{221801690880000} \zeta_2
\nonumber\\ && 
        -\frac{6736570381}{2000376000} l_2 \zeta_2
        +
        \frac{623204}{297675} l_2^2 \zeta_2
        -\frac{9668233}{9525600} \zeta_2^2
        +\frac{4449995721067}{1536288768000} \zeta_3
        -\frac{10711}{90720} \zeta_2 \zeta_3
\nonumber\\ && 
        -\frac{3293}{36288} \zeta_5
\Biggr]
+s^4 \Biggl[
        -\frac{1532491738898017906343}{200695848527462400000}
        +\frac{228437}{238140} a_4
        +\frac{228437}{5715360} l_2^4
\nonumber\\ && 
        -\frac{5429386088673075803}{1245638295982080000} \zeta_2
        +\frac{1669776162077}{422479411200} l_2 \zeta_2
        +\frac{1963141}{17463600} l_2^2 \zeta_2
        -\frac{7147363}{75442752} \zeta_2^2
\nonumber\\ && 
        +\frac{48280096513375243}{28316874571776000} \zeta_3
        +\frac{653}{45360} \zeta_2 \zeta_3
        -\frac{263}{9072} \zeta_5
\Biggr]
+s^5 \Biggl[
        -\frac{41000278323929714066450497}{16454972342427231191040000}
\nonumber\\ &&
        +\frac{1529009641}{2593344600} a_4
        +\frac{1529009641}{62240270400} l_2^4
        -\frac{6157090135300397592929177}{2142064386962175836160000} \zeta_2
\nonumber\\ && 
       +\frac{63096225504033647}{19936803414528000} l_2 \zeta_2
        +\frac{39051357271}{67426959600} l_2^2 \zeta_2
        -\frac{104412436391}{299675376000} \zeta_2^2
\nonumber\\ &&
        +\frac{23806175381893363619}{21537005745733632000} \zeta_3
        +\frac{23207}{1496880} \zeta_2 \zeta_3
        -\frac{977}{99792} \zeta_5
\Biggr]\Biggr\} 
 + {O}(s^6).
\end{eqnarray}

\section{Expansions at other points by analytic continuation}\label{Sec:AnalyticCont}

\vspace*{1mm}
\noindent
We can now use the expansions around $s=0$, for the differential equations 
of all form factors for each constant in these expansions, to compute the 
corresponding expansion at other points in $s$. Main targets are the 
expansions in the high energy limit $s \rightarrow \infty$, at the two-particle threshold 
($s=4$), 
and the four-particle pseudo-threshold ($s=16$). The expansions at other points are also 
important for two reasons:
\begin{enumerate}
  \item The calculation of these expansions are needed as intermediate steps to 
  reach these points.
  \item They are needed for accurate numerical evaluations of the form 
  factors in the whole kinematic range $-\infty < s < +\infty$.
\end{enumerate}
We will determine the coefficients of these expansions numerically with very high 
precision such that it will be sufficient to express the coefficients of the expansions 
around $s \rightarrow \infty$ in terms of special constants, as, e.g. multiple zeta values \cite{Blumlein:2009cf}
and others, using the PSLQ 
algorithm~\cite{PSLQ1,PSLQ2,Bailey:1999nv}, after a number of necessary matchings.

The radius of convergence of the expansions about $s=0$ depends on the constant 
in (\ref{eq:MZV-list}) under consideration, since the location of the nearest 
singularity is constant dependent. For the weight {\sf w}~=~4 constants, the 
closest singularity is located at $s=-1$, whereas for the remaining constants, 
it is at $s=-1/2$. To generate expansions about negative values of $s$, we may 
begin by choosing either a regular point between $s=0$ and the nearest singularity 
or the singular point itself as the center of the next expansion.

The basic idea of the method is to determine the coefficients of the expansion about 
the new point numerically by matching it to the expansion about $s=0$
within their common region of convergence. This procedure can then be iterated, constructing a sequence of expansions at progressively more negative values of $s$,
with each new expansion matched to the preceding one. The same strategy is applied for positive values of $s$.
In this way, the analytic continuation is carried out across the entire real axis, including through the singular points.
Once a sufficiently large positive or negative value of $s$ has been reached, 
the corresponding expansion can be matched to the asymptotic expansion at $s \rightarrow
\infty$. A corresponding calculation in Refs.~\cite{Fael:2022miw,Egner:2022jot,
Fael:2022rgm,Schonwald:2022djs,Fael:2023zqr,
Schonwald:2023uel} has been performed directly to the coupled systems of the master 
integrals by using the Frobenius method. For a general survey including existence results 
and also error bounds we refer, e.g., to~\cite{vdH:singhol,mezzarobba2016,AKR:22} and 
references therein.

\subsection{The structure of the local expansions at the different points of interest}

\vspace*{1mm}
\noindent
Since the poles in $\varepsilon$ of all form factors can be obtained in terms of known functions, the method we are about to describe will only be
applied to the ${O}(\varepsilon^0)$ terms. Therefore, we drop the index $i=0$ in the functions $f_{I;i,j,l}^{(0)}(s)$, and simply write $f_{I;j,l}^{(0)}(s)$.
Furthermore, we replace the label ``(0)'' by ``($s_0$)'' to indicate that the expansions are now about $s=s_0$ instead of $s=0$. 
For regular points, the expansions will have the following form

\begin{equation}
f_{I;j,l}^{(s_0)}(s) = \sum_{k=0}^{\infty} b_{I;j,l,k}^{(s_0)} z^k,
\label{eq:expansion.at.s=s0}
\end{equation}
where $z=s-s_0$.
For $s_0 \neq 0$, the coefficients $b_{I;j,l,k}^{(s_0)}$ will 
not necessarily be real, but also complex numbers. 

Around singular points, the expansions may include powers of logarithms and negative powers
 in $z$,
and the relation between $z$ and $s$ may involve square roots. All of this can be determined by studying the differential equations. 
In general, the expansions at the singular points, assuming that we are 
considering a constant $\kappa_l$ such that the point is indeed singular. A 
series of representations is

\begin{eqnarray}
f_{I;j,l}^{(-1/2)}(s) &=& \sum_{k=0}^{\infty} \sum_{i=0}^1
b_{I;j,l,i,k}^{(-1/2)} z^k \ln^i(z), \quad {\rm with} \quad z=s+1/2,
\label{eq:expansion.at.s=sm1h} \label{Equ:zExpansionStart}\\
f_{I;j,l}^{(-1)}(s) &=& \sum_{k=-3}^{\infty} b_{I;j,l,k}^{(-1)} z^k, \quad
{\rm with} \quad z=\sqrt{1+s}, \label{eq:expansion.at.s=sm1} \\
f_{I;j,l}^{(-4)}(s) &=& \sum_{k=-2}^{\infty} \sum_{i=0}^2
b_{I;j,l,i,k}^{(-4)} z^k \ln^i(z), \quad {\rm with} \quad z=s+4,
\label{eq:expansion.at.s=sm4} \\
f_{I;j,l}^{(1)}(s) &=& \sum_{k=-6}^{\infty} \sum_{i=0}^1
b_{I;j,l,i,k}^{(1)} z^k \ln^i(z), \quad {\rm with} \quad z=s-1,
\label{eq:expansion.at.s=s1} \\
f_{I;j,l}^{(3)}(s) &=& \sum_{k=-3}^{\infty} b_{I;j,l,k}^{(3)} z^k, \quad
{\rm with} \quad z=\sign{3-s} \sqrt{3-s},
\label{eq:expansion.at.s=s3} \\
f_{I;j,l}^{(4)}(s) &=& \sum_{k=-5}^{\infty} \sum_{i=0}^3
b_{I;j,l,i,k}^{(4)} z^k \ln^i(z), \quad {\rm with} \quad z=\sign{4-s}
\sqrt{4-s}. \label{eq:expansion.at.s=s4} \\
f_{I;j,l}^{(16)}(s) &=& \sum_{k=0}^{\infty}
b_{I;j,l,k}^{(16)} z^k, \quad {\rm with} \quad z= -i\sqrt{s-16}.
\label{eq:expansion.at.s=s16}
\end{eqnarray}
Likewise, the expansion at $s \rightarrow \infty$ is given by
\begin{eqnarray}
   f_{I;j,l}^{(\infty)}(s) = \sum_{k=-4}^{\infty} \sum_{i=0}^6
b_{I;j,l,i,k}^{(\infty)} z^k \ln^i(z), \quad {\rm with} \quad
z=\sqrt{-\frac{1}{s}}.
\label{eq:expansion.at.s=infinity}
\end{eqnarray}
In these equations, the lowest allowed values of $k$ and the
highest powers of the logarithms are determined by the differential
equations, considering all values of $j$ and $l$ for all form factors.
In some cases, the starting value of $k$ is higher; in such cases,
the corresponding coefficient $b_{I;j,l,i,k}^{(s_0)}$ is set to zero.

The full form factors can be written at ${O}(\varepsilon^0)$ in a similar way as in Eq.~(\ref{eq:F2fjl}),
\begin{equation}
  F_I^{(s_0)}(s) = \sum_{j=1}^3 C_j \sum_{l=1}^{10} \kappa_l f_{I;j,l}^{(s_0)}(s).
\label{eq:F2fjl-s0}
\end{equation}
After combining all terms in Eq.~(\ref{eq:F2fjl-s0}), all logarithms and square roots 
disappear around the spurious singular points $s=-1/2, -1, -4, 1, 3$, and we are left 
with ordinary expansions about these points like the ones 
in Eq.~(\ref{eq:expansion.at.s=s0}). In the case of the expansions at infinity, the 
logarithms remain, but the square roots disappear, and we end up with 
an asymptotic expansion in $-1/s$, see Section~\ref{Sec:HighEnergyLimit}.

\begin{center}
\begin{figure}[H]
\begin{center}
     \includegraphics[width=0.50\textwidth]{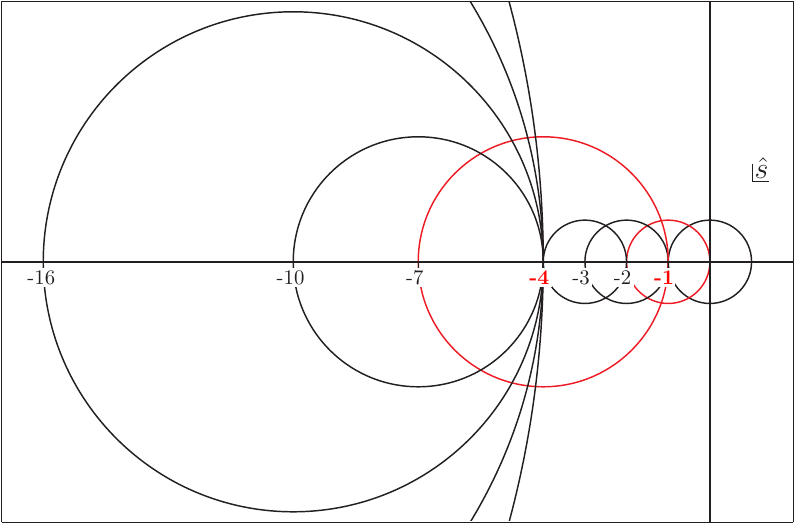}
\end{center}
\caption{\sf Radii of convergence of the successive points around which expansions are matched in the case of the weight {\sf w}~=~4 constants for $s<0$. 
The points $s=-1$ and $s=-4$, shown in red, are singularities of the differential equations, which disappear after everything is combined in Eq.~(\ref{eq:F2fjl-s0}).} 
\label{fig:convergence1a}
\end{figure}
\end{center}

The set of expansion points must include the singular points of the differential equations.
Beyond this, there is considerable freedom in the choice of regular expansion points.
For the weight {\sf w}~=~4 constants in  (\ref{eq:MZV-list}), and for $s<0$, we chose the following set of expansion points,
\begin{equation}
  s_0 \in \left\{-196, \, -100, \, -52, \, -28, \, -16, \, -10, \, -7, \, -4, \, -3, \, -2, \, -1\right\}.
\label{eq:s0<0-w4}
\end{equation}
This choice is motivated by the radii of convergence associated with the different expansion points. Since the expansions are matched sequentially,
each successive expansion point in the negative direction is chosen to lie on the boundary of the radius of convergence of the preceding one,
as illustrated in Figure~\ref{fig:convergence1a}.
This strategy maximizes the overlap between the regions of convergence of consecutive expansions, thereby facilitating the matching procedure.
For example, since the radius of convergence of the expansions at $s=-4$ is equal to 3 (the nearest singularity is at $s=-1$),
the next expansion point is chosen at $s=-7$. The expansion at $s=-7$  also has a radius of convergence equal to 3 (nearest singularity at $s=-4$), so the next expansion
is at $s=-10$, which has a radius of convergence equal to 6, so the next expansion is chosen at $s=-16$.

Similar illustrations apply to the convergence circles of the other expansion points in $s$.
For $s>0$, we chose
\begin{equation}
  s_0 \in \left\{1, \, 2, \, 3, \, 4, \, 5, \, 6, \, 8, \, 12, \, 16 \right\}.
\label{eq:s0>0-w4}
\end{equation}
The expansions about $s=1, 2, 3$ were constructed sequentially to reach 
the two-particle threshold at $s=4$.

The expansion about $s=16$, corresponding to the four-particle pseudo-threshold, 
was instead obtained by matching directly to the expansion around $s \rightarrow \infty$.
Consequently, the intermediate expansion points at $s= 5, 6, 8, 12$, are not 
strictly required for this construction.
Nevertheless, we computed these additional expansions to enable an efficient 
numerical evaluation of the form factors with very high precision over the entire kinematic region.

In the case of the constant $\kappa_4 = l_2 \zeta_2$, we also have singularities about $s=-1/2$ and $s=-1$.
Therefore, for this constant we chose the following expansion points for $s<0$,
\begin{equation}
  s_0 \in \left\{-196, \, -100, \, -52, \, -28, \, -16, \, -10, \, -7, \, -4, \, -3, \, -2, \, -\frac{3}{2}, \,  -1, \, -\frac{1}{2} \right\}.
\label{eq:s0<0}  
\end{equation}

For $s>0$, we adopted a slightly different strategy from that based solely on the radii of convergence.
Instead, we performed expansions around every integer and half-integer value of $s$ up to $s=4$. The complete set of expansion points is
\begin{equation}
  s_0 \in \left\{\frac{1}{2}, \, 1, \, \frac{3}{2}, \, 2, \, \frac{5}{2}, \, 3, \, \frac{7}{2}, \, 4, \, 5, \, 6, \, 8, \, 12, \, 16 \right\}.
\label{eq:s0>0}
\end{equation}

The inclusion of these additional intermediate expansion points improves the numerical precision of the matching procedure,
since shorter matching steps reduce the accumulation of numerical errors.

In the case of the constants $\kappa_1 = 1$, $\kappa_2 = \zeta_2$ and $\kappa_3 = \zeta_3$,
the expansion points in (\ref{eq:s0<0}) and (\ref{eq:s0>0}) were not enough to achieve sufficiently high precision.
Therefore, in the case of these constants, we included expansion points in middle of each pair of points in (\ref{eq:s0<0})
and each pair of points in (\ref{eq:s0>0}) up to $s=4$, since, as we already mentioned in the case of $l_2 \zeta_2$, the more points we include,
the higher the precision that can be achieved. For $s<0$, we then have
\begin{eqnarray}
  s_0 &\in& \left\{ -196, \, -148, \, -100, \, -76, \, -52, \, -40, \, -28, \, -22, \, -16, \, -13, \, -10, \, -\frac{17}{2}, \, -7, \, -\frac{11}{2}, \right. \nonumber \\
         && \left. \,\,\, -4, \, -\frac{7}{2}, \, -3, \, -\frac{5}{2}, \, -2, \, -\frac{7}{4}, \, -\frac{3}{2}, \, -\frac{5}{4}, \,  -1,
                   \, -\frac{3}{4}, \, -\frac{1}{2}, \, -\frac{1}{4} \right\},
\label{eq:s0<0-full}
\end{eqnarray}
while for $s>0$, we chose
\begin{equation}
  s_0 \in \left\{\frac{1}{4}, \, \frac{1}{2}, \, \frac{3}{4}, \,  1, \, \frac{5}{4}, \, \frac{3}{2}, \, \frac{7}{4}, \, 2, \, \frac{9}{4},
                   \, \frac{5}{2}, \, \frac{11}{4}, \, 3, \, \frac{13}{4}, \, \frac{7}{2}, \, \frac{15}{4}, \, 4, \, 5, \,  6, \, 8, \, 12, \, 16 \right\}.
\label{eq:s0>0-full}
\end{equation}

It is worth noting that, in the case of the constants $\kappa_1 = 1$ and $\kappa_2 = \zeta_2$, the differential equations exhibit a singularity at $s=16/3$.
To avoid this singular point during the matching procedure, we obtain the expansions around $s=5$ by matching them directly to the expansions around $s=4$.
The expansions around $s=6$, $s=8$, and $s=12$ are instead obtained in the reverse direction:
the expansions around $s=16$ (which can themselves be determined by matching to the 
expansions at $s \rightarrow \infty$) are matched to the expansions around $s=12$,
which are subsequently matched to those around $s=8$ and $s=6$.

\subsection{Computing the expansion at the individual points}

\vspace*{1mm}
\noindent
When carrying out the analytic continuation starting at $s=0$, 
it was necessary to compute a basis of the truncated series solutions at the 
individual points up to a high order $N$.
In the following, we focus on variants of the computation implemented in the new 
{\tt Mathematica}
package \texttt{HolonomicContinuation}\footnote{The freely available package \texttt{HolonomicContinuation}
can be downloaded from \url{https://risc.jku.at/sw/HolonomicContinuation/}.} with a 
special focus on computer algebra tools to deal with huge differential equations. 

The differential equations in $s$ can be transformed into differential equations 
in $z = s - s_0$, both for singular
and regular points (see~\eqref{eq:expansion.at.s=s0}--\eqref{eq:expansion.at.s=infinity}), 
using the corresponding change of variables
to get a differential equation
\begin{equation}\label{Equ:DEForPoint}
	a_0(z)F(z)+a_1(z)\frac{d}{dz} F(z)+\dots+a_r(z)\frac{d^r}{dz^r}F(z)=0
\end{equation}
with polynomials $a_i(z)$ in $z$ and rational number coefficients. 
The general expansion in $z$ for the form factors have then the following shape:
\begin{equation}\label{Equ:AnsatzLogPowerSeriesGeneral}
	F(z)=\sum_{i=0}^k\ln^i(z)\sum_{j=l}^{\infty} c_i(j) z^j \text{ with } l\in\mathbb Z\text{ and }c_i(j)\in \mathbb Q.
\end{equation}

\subsubsection{Log-free contributions}

\vspace*{1mm}
\noindent
In the case of regular points, substituting the corresponding series 
expansion~\eqref{eq:expansion.at.s=s0} with $z=s-s_0$ 
into these differential equations 
yields relations
among the coefficients $b_{I;j,l,k}^{(s_0)}$.
As a result, all coefficients can be expressed in terms of a finite set of independent coefficients, whose number is equal to the order of the differential equation.

As an illustrative example, consider the pseudoscalar function $f_{p;3,7}^{(-2)}(s)$, corresponding to the color factor $C_3=C_A^2 C_F$ and the constant $\kappa_7=l_2^4$.
Its expansion is given by Eq.~(\ref{eq:expansion.at.s=s0}), with $s_0=-2$.
Substituting this expansion into the corresponding differential equation in $z$, 
which is of order 9, see Table~\ref{orderstable},
reveals that the first 9 coefficients,
\begin{equation}
  \left\{ b_{p;3,7,0}^{(-2)} \, , \, b_{p;3,7,1}^{(-2)} \, , \cdots \, , b_{p;3,7,8}^{(-2)}\right\},
\label{eq:free.coeffs.sm2}
\end{equation}
form a set of independent coefficients. All subsequent coefficients 
are expressed as linear combinations of those  over
explicitly given rational numbers.

We can similarly obtain relations between the coefficients when the change of variables from $s$ to $z$ involves a square root.
In the log-free case, this can occur only at the singular points $s=-1$, $s=3$, and $s=16$,
see Eqs.~(\ref{eq:expansion.at.s=sm1}), (\ref{eq:expansion.at.s=s3}) and (\ref{eq:expansion.at.s=s16}).
For example, consider the pseudoscalar function $f_{p;1,5}^{(-1)}(s)$,
corresponding to the color factor $C_1=C_F^3$ and the constant $\kappa_5=\zeta_2^2$.
Its expansion is given by Eq.~(\ref{eq:expansion.at.s=sm1}).
Substituting this expansion into the corresponding differential equation 
in $z$, which is of order 13, cf.~Table~\ref{orderstable},
one finds that every coefficient in the series can be expressed as a rational linear combination of the following 13 independent coefficients:
\begin{eqnarray}
	&& \left\{ b_{p;1,5,-1}^{(-1)} \, , \, b_{p;1,5,0}^{(-1)} \, , \, b_{p;1,5,2}^{(-1)} \, , \, b_{p;1,5,4}^{(-1)} \, , \, b_{p;1,5,6}^{(-1)} \, , \, b_{p;1,5,8}^{(-1)} \, , \,
	b_{p;1,5,10}^{(-1)} \, , \right. \nonumber \\
	&& \left.  \,\,\,\, b_{p;1,5,12}^{(-1)} \, , \,b_{p;1,5,14}^{(-1)} \, , \,b_{p;1,5,16}^{(-1)} \, , 
\,b_{p;1,5,18}^{(-1)} \, , \,b_{p;1,5,20}^{(-1)} \, , \, b_{p;1,5,22}^{(-1)} \right\}.
	\label{eq:free.coeffs.sm1}
\end{eqnarray}
Note that the coefficients of $z$ with odd powers depend only on $b_{p;1,5,-1}^{(-1)}$. 
One has
\begin{eqnarray}
	b_{p;1,5,1}^{(-1)} &=& \frac{9}{40} b_{p;1,5,-1}^{(-1)} \\
	b_{p;1,5,3}^{(-1)} &=& \frac{14489}{960} b_{p;1,5,-1}^{(-1)} \\
	b_{p;1,5,5}^{(-1)} &=& \frac{55417}{25600} b_{p;1,5,-1}^{(-1)} \\
	\vdots      &=&      \vdots \nonumber
\end{eqnarray}
The coefficients multiplying even powers of $z$, on the other hand,
are rational linear combinations of the independent even-power coefficients 
listed in Eq.~(\ref{eq:free.coeffs.sm1}).
The rational coefficients rapidly acquire extremely large numerators and 
denominators.
Their size increases significantly for coefficients of higher powers, making exact symbolic manipulations increasingly demanding.
This approach remains feasible in cases where the number of coefficient relations required to achieve the desired numerical precision is not too large,
such as for the weight {\sf w}~=~4 constants and the $l_2\zeta_2$ contribution.
For the cases $\kappa_1=1$, $\kappa_2=\zeta_2$, and $\kappa_3=\zeta_3$, however,
substantially more coefficient relations are needed, making the direct symbolic approach impractical. 

We consider here the case~\eqref{Equ:AnsatzLogPowerSeriesGeneral} with $k=0$ and $l=0$, i.e., for log-free series expansions of the form 
\begin{equation}\label{Equ:AnsatzPowerSeries}
	F(z)=\sum_{j=0}^{\infty} c(j) z^j \text{ with }c(j)\in \mathbb Q.
\end{equation}
We employ an improved procedure to determine $r$ linearly independent (truncated) power 
series solutions (which can be expressed by $r$ free parameters).
Plugging~\eqref{Equ:AnsatzPowerSeries} into~\eqref{Equ:DEForPoint} and performing coefficient comparison with respect to\ $z^j$ yields a recurrence for the coefficients $c(j)$ of
the series solutions of the form
\begin{equation}\label{Equ:RecForOrdinaryPoint}
	b_0(j)c(j)+b_1(j)c(j+1)+\dots+b_u(j)c(j+u)=0
\end{equation}
for some polynomials $b_i(j)$ in $j$.

For the differential equation~\eqref{Equ:DEForPoint} of order $r$ coming from the substitution $s-s_0\to z$, it suffices to give $F(z)$ up to order $r-1$ to uniquely
determine the full solution $F(z)$. To get a basis of all truncated
series solutions we set as initial terms $F_i(z)=\sum_{j=0}^{r-1} \delta_i(j)z^j$ for $0\le i<r$, where $\delta_i(j)$ is $1$ if $j=i$ and $0$ otherwise.
By using the recurrence~\eqref{Equ:RecForOrdinaryPoint}, we prolong $F_i(s)$ for each $i$ up to $z^N$ for some large $N\in \mathbb{N}$.
At first glance, if the order $u$ of the recurrence is larger than $r$, more initial terms are needed to get a similar statement for the sequences which are a solution
of the recurrence. However, note that the sequence $c(j)$ can be trivially extended to 
$-\infty$ be setting $c(j)=0$ for $j<0$.
Furthermore, by Theorem~3.5, Ref.~\cite{Kaue2023}, the full two-sided sequence 
$(c(j))_{j\in\mathbb Z}$ is a solution of the recurrence.
That is, we extend the initial values $c(0),\ldots, c(r-1)$ to $c(j)$ for $u-r\le j<r$ and can compute $c(r),\ldots, c(N)$ using the recurrence.

For the substitution $\sqrt{s-s_0}\to z$ one needs to split the problem between the even and the odd cases accordingly.
As shown in the example given above for~(\ref{eq:expansion.at.s=sm1}),
we determine the free coefficients in advance by naive linear system solving and proceed accordingly to obtain $r$ linearly independent (truncated) power series solutions in $z$
as described previously.

For the constant $\kappa_1=1$, carrying out this computation in reasonable time was especially challenging for the following reasons:
\begin{enumerate}
      \item The coefficients of the polynomials in the differential equations are very large, sometimes more than 10000 digits.
      \item The order and the degree of the differential 
equations are very large, up to $102$ and $4078$, 
	  respectively, see Tables~\ref{orderstable} and~\ref{degreestable}. 
Therefore, 
also the order and the degree of the associated recurrences,
          which are obtained using closure properties, are very large, up to 4078 and 102. 
      \item The obtained recurrences are numerically ill-conditioned. That is, even if very precise approximations for $c(k),\ldots, c(k+u-1)$ are supplied to the recurrence,
        the computed values $\tilde c(k+u),\tilde c(k+u+1),\ldots$ lose precision very quickly.
\end{enumerate}
Note that it is not necessary to compute the exact values 
$c(0),\ldots, c(N)\in \mathbb Q$. Precise approximations are sufficient.
However, due to the third reason mentioned above, doing the computation 
numerically was not feasible, and the entire calculation of the truncated 
series solutions had to be performed using exact numbers. 

The enormous size of the recurrences made the computation using rational numbers very expensive, which is why we chose a different approach.
Instead of directly computing the $c(j)\in \mathbb{Q}$, we first compute $c(j)$ modulo several (machine-sized) primes.
Then we use the Chinese Remainder Theorem and rational number reconstruction~\cite{vonzurgathen2013}, to retrieve the exact $c(j)\in \mathbb{Q}$.
For maximal performance, we implemented this in C using the C-library {\tt FLINT}~\cite{flint}.
This approach turned out to be more than 25 times\footnote{Here we compared with our first implementation in 
{\tt FLINT},
which was based on pure rational number arithmetic (using the same algorithm) 
for $C_1$ and $\kappa_1=1$, i.e., $f_{s;1,1}(s)$ for 15000 values.
($4$ vs.\ $101$ CPU-hours)} faster than our best implementation based on pure rational number arithmetic.
For an efficient implementation, the following three observations were crucial:
\begin{enumerate}
\item Since the computation time in this approach is dominated by the evaluation of the polynomials of the recurrence,
  it is important to handle all $r$ elements in the basis of truncated series solutions at the same time.
  In this way the polynomials have to be evaluated only once for each prime.
\item The denominator of consecutive values $\operatorname{den}(c(j))$ and $\operatorname{den}(c(j+1))$ are very similar,
  that is, the quotient of these denominators is a rational number of small size.
  So for reconstructing $c(j+1)$, one can multiply the computed modular images $c(j+1)\mod p$ with $\operatorname{den}(c(j))$
  and then reconstruct $\operatorname{den}(c(j))\,c(j+1)$. With this approach, only about half as many primes are needed,
  since we observed that in our case $\operatorname{den}(c(j))c(j+1)$ has about half the size (half the number of digits) as $c(j+1)$.
\item Note that the size of $c(j)$ increases approximately linearly in $j$, so if $j_1<j_2$, then less primes are sufficient for
  reconstructing $c(j_1)$ than for $c(j_2)$. If the reconstruction succeeded already for 
$c(M-u+1),\ldots, c(M)\in \mathbb Q$ for some $M>u$,
  then these values can be used as new initial values for 
next primes to compute $c(M+1),c(M+2),$ etc.
\end{enumerate}
To save memory, the final result is transformed from exact rational numbers to a numerical representations with up to 5000 digits.

The above approach for power series expansions~\eqref{Equ:AnsatzPowerSeries} in $z$ can be extended straightforwardly to Laurent series expansions with $l\in\mathbb Z$.

\subsubsection{Expansions with log-contributions}

\vspace*{1mm}
\noindent
Finally, we consider the general case of our form factor calculation having expansions of 
the form~\eqref{Equ:AnsatzLogPowerSeriesGeneral} with $k \geq 1$.
The most involved situation appeared with $k=6$ at the point $
s \rightarrow \infty$. For simplicity, we 
restrict~\eqref{Equ:AnsatzLogPowerSeriesGeneral} to $l=0$.
For this particular case, we need to compute a basis of the truncated versions of solutions of the form 
\begin{equation}\label{Equ:AnsatzLogPowerSeries}
	F(z)=\sum_{i=0}^k\ln^i(z)\sum_{j=0}^{\infty} c_i(j) z^j \text{ and }c_i(j)\in \mathbb Q.
\end{equation}
As in the log-free situation, exactly $r$ values $c_i(j)$ uniquely determine all the other coefficients; however, they distribute over the different log-contributions.

To handle this situation, we determine in a preprocessing step the initial values by plugging in the ansatz~\eqref{Equ:AnsatzLogPowerSeries},
matching coefficients for $\ln^i(z)z^j$ for $i=0,\dots,k$ and $j=0\dots, \rho$ and 
deriving $(k+1)\rho$ constraints in $(k+1)\rho$
unknowns; if $s-s_0\to z$ we can take $\rho=r$, and if $\sqrt{s-s_0}\to z$ we take 
$\rho=2r$.\footnote{Usually, we set $\rho$ to be larger than the order of the given
differential equation to handle the pole situation arising from the underlying recurrence.} 
Solving for this system enables one to express the  $(k+1)\rho$ coefficients by a linear combination of $r$ unknowns.
In the generic case, we found  that the $c_i(j)$ for $0\leq j<n_i$ with $n_i\geq1$ and $n_0+\dots+n_k=\rho$ remain free and the other coefficients are determined linearly by them.

\noindent\textit{Remark.} In the largest case (the constant case $\kappa_1$ with expansions 
around $\infty$), the order of the differential equations was $r=102$ and $k=6$.
Due to technical reasons we had to set even $r=140$, i.e., had to solve linear systems of sizes around $1000\times1000$.
To perform this efficiently, the solutions of the linear system has been obtained by solving the system 
modulo several (machine-sized) primes, using the Chinese Remainder Theorem and rational number reconstruction.

With this data, we proceeded as follows to compute the first $N+1$ coefficients of the $r$ linearly independent solutions of the form~\eqref{Equ:AnsatzLogPowerSeries}
for~\eqref{Equ:DEForPoint} efficiently.
Plugging the ansatz~\eqref{Equ:AnsatzLogPowerSeries} into~\eqref{Equ:DEForPoint} and performing coefficient comparison
with respect to $\ln^k(z)z^j$ yields~\eqref{Equ:RecForOrdinaryPoint}
with $c(j)=c_k(j)$.
Using our earlier calculation to determine the initial values, we obtain $n_k$ linearly independent truncated sequences of length $r$.
Reusing the technology for the log-free contribution we can now compute the first $N+1$ coefficients efficiently of the $\ln^k(z)$-contribution in terms
of the free coefficients $c_k(0),\dots,c_k(n_k-1)$. 
Now define
\begin{eqnarray}
G(z)=\ln^k(z)\sum_{j=0}^{\infty} c_k(j)z^j
\end{eqnarray}
and
\begin{eqnarray}
\tilde{F}(z)=\sum_{i=0}^{k-1}\ln^i(z)\sum_{j=0}^{\infty} c_i(j) z^j,
\end{eqnarray}
i.e., $F(z)=\tilde{F}(z)+G(z)$. Thus replacing $F(z)$ by 
$\tilde{F}(z)+G(z)$ in~\eqref{Equ:DEForPoint} and moving $G(z)$ (and their derivatives) 
to the right hand side leads to
\begin{eqnarray}
a_0(z)F(z)+a_1(z)\frac{d}{dz} F(z)+\dots+a_r(z)\frac{d^r}{dz^r}F(z)=H(z)
\end{eqnarray}
with
\begin{eqnarray}
H(z)=\sum_{i=0}^{k-1}\ln^i(z)\sum_{j=-v}^{\infty} d_i(j) z^j,
\end{eqnarray}
where the $d_i(j)\in\mathbb{Q}$ up to the order $N$ in $z$ are given explicitly. Now we repeat this strategy. By plugging in $\tilde{F}(z)$
into this new differential equation and doing coefficient comparison with respect to\ $\ln^{k-1}(z) z^j$ yields a linear recurrence of the form
\begin{equation}\label{Equ:RecInhomForLog}
	b_0(j)c_{k-1}(j)+b_1(j)c_{k-1}(j+1)+\dots+b_u(j)c_{k-1}(j+u)=d_{k-1}(j+\ell)
\end{equation}
for some integer $\ell$. Again we can use the knowledge of the initial values coming from the preprocessing step to prolong the values for $j=r,\dots,N$
by using the inhomogeneous recurrence. Here we can again apply the tools we used for the log-free case with a slight modification to handle also inhomogeneous contributions.
Applying this method iteratively, for $k,k-1,\dots,0$ delivers the desired $r$ solutions of~\eqref{Equ:AnsatzLogPowerSeries} up to the order $N$.
In this process we remark that the left hand-side of~\eqref{Equ:RecInhomForLog} remains the same and only the right-hand sides need to be updated.
This observation can be utilized to produce only once~\eqref{Equ:RecInhomForLog} for a symbolic (unevaluated) right-hand side that subsequently can
be used to produce the new explicit right-hand sides for $k,k-1,\dots,0$. 

The above approach for power series expansions~\eqref{Equ:AnsatzPowerSeries} in $z$ can be extended straightforwardly to Laurent series expansions with $l\in\mathbb Z$.
As an example, consider the axial-vector function $f_{a,2 \, ; \, 3,3}^{(-4)}(s)$, corresponding to the color factor $C_3=C_A^2C_F$ and the constant $\kappa_3=\zeta_3$.
Its expansion is given by Eq.~(\ref{eq:expansion.at.s=sm4})\footnote{In this particular
 case, the lower bound of the summation index $k$ is $-2$, instead of $-4$.
	As we explained before, the corresponding coefficients for $k < -2$ are then 
equal to zero.}.
The corresponding differential equation in $z$ is of order 49, see 
Table~\ref{orderstable}.
Every coefficient in the expansion can be expressed as a linear combination of the following 49 independent coefficients:
\begin{eqnarray}
   &&\left\{b_{a,2 \, ; \, 3,3,0,-2}^{(-4)} \, , \, b_{a,2 \, ; \, 3,3,0,0}^{(-4)} \, , \, b_ {a,2 \, ; \, 3,3,0,1}^{(-4)} \, , b_{a,2 \, ; \, 3,3,0,2}^{(-4)} \, , \ldots, b_{a,2 \, ; \, 3,3,0,43}^{(-4)} \, ,
	\right. \nonumber \\
   && \left. \,\,\,\, b_{a,2 \, ; \, 3,3,1,0}^{(-4)} \, , \, b_{a,2 \, ; \, 3,3,1,1}^{(-4)} \, , \, b_{a,2 \, ; \, 3,3,2,0}^{(-4)} \, , \, b_{a,2 \, ; \, 3,3,2,1}^{(-4)} \right\}.
\label{eq:free.coeffs.sm4}
\end{eqnarray}
The following are a few illustrative examples of the relations between the coefficients:
\begin{eqnarray}
	b_{a,2 \, ; \, 3,3,2,2}^{(-4)} &=& 0.007812500000\, b_{a,2 \, ; \, 3,3,2,0}^{(-4)} - 0.03125000000\, b_{a,2 \, ; \, 3,3,2,1}^{(-4)} \\
	b_{a,2 \, ; \, 3,3,2,3}^{(-4)} &=& -0.03356933594\, b_{a,2 \, ; \, 3,3,2,0}^{(-4)} - 0.07031250000\, b_{a,2 \, ; \, 3,3,2,1}^{(-4)} \\
	b_{a,2 \, ; \, 3,3,1,2}^{(-4)} &=& 0.001815795898\, b_{a,2 \, ; \, 3,3,0,-2}^{(-4)} + 0.007812500000\, b_{a,2 \, ; \, 3,3,1,0}^{(-4)} \nonumber \\
	&& - 0.03125000000\, b_{a,2 \, ; \, 3,3,1,1}^{(-4)} + 4.121093750\, b_{a,2 \, ; \, 3,3,2,0}^{(-4)} \nonumber \\
	&& + 9.000000000\, b_{a,2 \, ; \, 3,3,2,1}^{(-4)} \\
	b_{a,2 \, ; \, 3,3,1,3}^{(-4)} &=& 0.001866340637\, b_{a,2 \, ; \, 3,3,0,-2}^{(-4)} - 0.03356933594\, b_{a,2 \, ; \, 3,3,1,0}^{(-4)} \nonumber \\
	&& - 0.07031250000\, b_{a,2 \, ; \, 3,3,1,1}^{(-4)} + 2.156548394\, b_{a,2 \, ; \, 3,3,2,0} \nonumber \\
	&& + 4.407118056\, b_{a,2 \, ; \, 3,3,2,1}^{(-4)} \\
	b_{a,2 \, ; \, 3,3,0,44}^{(-4)} &=& 1.295717677 \times 10^{-59}\, b_{a,2 \, ; \, 3,3,0,-2}^{(-4)} - 1.460958567 \times 10^{-48}\, b_{a,2 \, ; \, 3,3,0,2}^{(-4)} \nonumber \\
	&& + 2.564253656 \times 10^{-44}\, b_{a,2 \, ; \, 3,3,0,3}^{(-4)}+ \ldots + 5.652741729\, b_{a,2 \, ; \, 3,3,0,43}^{(-4)} \nonumber \\
	&& - 4.777417126 \times 10^{-58}\, b_{a,2 \, ; \, 3,3,1,0}^{(-4)} - 8.784518590 \times 10^{-58}\, b_{a,2 \, ; \, 3,3,1,1}^{(-4)} \nonumber \\
	&& + 1.459368840 \times 10^{-51}\, b_{a,2 \, ; \, 3,3,2,0}^{(-4)} + 2.847550234 \times 10^{-51}\, b_{a,2 \, ; \, 3,3,2,1}^{(-4)}\, .
\end{eqnarray}
The floating point numbers in front of the coefficients were computed with 5000 digits. 
Here we are showing only 10 of them.

\subsection{Determining the free coefficients by matching}

\vspace*{1mm}
\noindent
The coefficient relations can now be substituted into large truncated versions of the corresponding series expansions,
i.e., expansions in which the sums over $k$ in Eqs.~(\ref{eq:expansion.at.s=s0}--\ref{eq:expansion.at.s=infinity}) are truncated at a sufficiently large value $N$.
In this way, each truncated expansion is expressed solely in terms of its independent coefficients.
The expansions can then be evaluated at any point within their respective radii of convergence,
allowing consecutive expansions with overlapping regions of convergence to be matched and the independent coefficients to be determined numerically.

To estimate the numerical precision of the coefficients obtained in this way,
we introduce an additional coefficient into each expansion corresponding to a term that is known to be absent.
For example, the expansion around a regular point $s_0$ contains only non-negative powers of $z$.
We therefore formally extend the expansion to start at $k=-1$, introducing the coefficient $b_{I;j,l,i,-1}^{(s_0)}$.
Since this coefficient is known to vanish identically, any non-zero value obtained from the matching procedure provides a direct estimate of the numerical error.
We refer to this quantity as the {\itshape test coefficient}.

Suppose that the independent coefficients of one expansion have already been determined numerically.
To determine the coefficients of the next expansion, written in terms of its independent coefficients together with the test coefficient,
we evaluate both expansions at $r+1$ distinct points in the overlap of their convergence regions\footnote{The evaluation is performed using the Feynman prescription,
according to which $s$ is understood to have an infinitesimal positive imaginary part. This is particularly important for expansions around singular points,
where logarithms and square roots are present.} and equate the corresponding values.
This yields a linear system of $r+1$ equations for the $r+1$ unknown coefficients. A convenient choice of matching points is
\begin{equation}
\xi \in \left\{s_{\rm ini}+n \delta; \,\, n=0,\ldots,r \right\},
\label{eq:matching.points}
\end{equation}
where the initial point $s_{\rm ini}$ and the spacing $\delta$ are chosen so that every point 
in $\xi$ lies within the common region of convergence of the two expansions. The values of $\delta$ were 
chosen in the range from $1/5000$ to $1/20000$.

The numerical accuracy of the matching depends on the choice of the points $\xi$.
In practice, we scan different values of $s_{\rm ini}$ and $\delta$ and select the combination that minimizes the magnitude of the test coefficient,
thereby maximizing the precision of the extracted independent coefficients.

Once the independent coefficients have been determined numerically, they can be substituted into the corresponding coefficient relations to obtain
all remaining coefficients numerically. This, in turn, allows the corresponding expansion to be evaluated numerically at any value of $s$ within its radius of convergence.

The precision of the coefficients obtained with the matching procedure also depends on the truncation order $N$.
As expected, increasing $N$ generally improves the numerical accuracy, but at the cost of a more demanding computation.
The value of $N$ must therefore be chosen large enough to achieve the desired precision, while keeping the computational cost within reasonable limits.

As mentioned previously, the expansion around $s=0$ was computed up to 
$N=100000$ terms.
Although such a large truncation order is not strictly necessary, it could be obtained efficiently using the corresponding recurrence relations,
requiring only moderate computational resources.
Moreover, the resulting highly accurate expansion provides an excellent starting point for the subsequent
matching procedure.

For the remaining expansion points, where the coefficients were obtained using the matching method described above, substantially smaller values of $N$ were sufficient.
For the weight {\sf w}~=~4 constants listed in (\ref{eq:MZV-list}), we used $N=5000$ for all expansion points with negative values of $s$ in (\ref{eq:s0<0-w4}),
while for the positive expansion points in (\ref{eq:s0>0-w4}) we used 
$N=10000$, except at $s=16$, where $N=5000$ was sufficient. At $s \rightarrow \infty$, 
we 
used $N=4000$.
These truncation orders were more than adequate to obtain the numerical precision required to reconstruct all coefficients of the asymptotic expansions
at $s \rightarrow \infty$ by applying the PSLQ algorithm. In practice, considerably 
smaller values 
of $N$ would likely have been sufficient.
However, since it is difficult to predict in advance the minimum truncation order required for a successful PSLQ reconstruction,
we opted for a large value that remained computationally feasible.
For the contribution proportional to $\kappa_4=l_2\zeta_2$, we used $N=5000$ for all finite expansion points in (\ref{eq:s0>0}) and (\ref{eq:s0<0}),
and $N=3000$ at $s \rightarrow \infty$.

The most demanding cases are those corresponding to $\kappa_1=1$, $\kappa_2=\zeta_2$, and $\kappa_3=\zeta_3$, for which substantially larger truncation orders are required.
For the expansion points listed in (\ref{eq:s0<0-full}), as well as for the 
expansions at $s \rightarrow \infty$, we used $N=10000$. 
For $\kappa_1=1$ and $s=-1$, $s=-5/4$, $s=-3/4$ we used even $N=16000$ coefficients. 
Computing expansions of this length is particularly challenging because of the high order of the differential equations
and the large degree of the associated polynomial coefficients, see 
Tables~\ref{orderstable} and \ref{degreestable}.
Such a large value of $N$ proved essential in the case $\kappa_1=1$ to achieve the precision required for the PSLQ reconstruction.
For the cases $\kappa_2=\zeta_2$ and $\kappa_3=\zeta_3$, somewhat smaller truncation orders, such as $N=8000$ or even $N=7000$,
would probably have been sufficient. Nevertheless, as in the weight {\sf w}~=~4 case, it was not possible to determine this beforehand,
and therefore we preferred to use a larger value to ensure that the numerical precision 
at $s \rightarrow \infty$ was comfortably adequate to use the PSLQ algorithm, 
see Section~\ref{Sec:HighEnergyLimit}.

For the positive expansion points listed in (\ref{eq:s0>0-full}), we used $N=8000$ for $\kappa_1=1$ and $N=4000$ for both $\kappa_2=\zeta_2$ and $\kappa_3=\zeta_3$.
Smaller values would probably have sufficed, since, unlike in~\cite{{Blumlein:2023uuq}},
our objective here is not necessarily to apply the PSLQ algorithm to the coefficients of the expansions around $s=4$.
Nevertheless, we again chose relatively large truncation orders in order to obtain a comparable level of numerical precision across all expansion points.
\section{The high energy limit} 
\label{Sec:HighEnergyLimit}

\vspace*{1mm}
\noindent
We consider the range $s \in [0, -\infty[$.
As we explained in Section \ref{Sec:AnalyticCont}, in order to approach the high energy limit we construct overlapping series moving from $s=0$ to 
$s=-\infty$. The (logarithmic modulated) expansions at $s \rightarrow \infty$ associated 
to each 
constant in Eq.~(\ref{eq:kappa-const}) and each color factor in Eq.~(\ref{eq:color-factors})
are given in Eq.~(\ref{eq:expansion.at.s=infinity}). We obtain the form factors by combining these expansions using Eq.~(\ref{eq:F2fjl-s0}),
after which the square roots disappear and we are left with integer powers of $-1/s$. The 
representation of the form factors around $s \rightarrow \infty$ is then given by
\begin{equation}
\label{eq:x0}
F_I(s) = \sum_{i=1}^3 C_i \sum_{l=0}^6 \sum_{k=0}^\infty {\tilde c}_{I;i,l,k} 
\ln^l\left(-\frac{1}{s}\right) 
\frac{1}{s^k}.
\end{equation}
A central question concerns the functional structure of the coefficients 
${\tilde c}_{I;i,l,k}$. In the quarkonic case they were given for each color factor by 
$\zeta$-values only.

After performing the matchings described in Section~\ref{Sec:AnalyticCont} we rationalized the
coefficients in Eq.~(\ref{eq:expansion.at.s=infinity}). Our accuracy has been chosen such that we could use
integer formalisms like
PSLQ \cite{PSLQ1,PSLQ2,Bailey:1999nv} to determine the corresponding coefficients in analytic form.
After summing all the contributions in Eq.~(\ref{eq:F2fjl-s0}),
the coefficients ${\tilde c}_{I;i,l,k}$ turned out to be rational combinations of the following set of constants,

\begin{eqnarray}
\label{eq:const}
&& \left\{
1,\,
\zeta_2,\,
\zeta_3,\,
\zeta_2^2,\,
\zeta_5,\,
\zeta_2^3,\,
\zeta_3^2,\,
\zeta_2 \zeta_3,\,
l_2^4,\,
l_2^5,\,
l_2 \zeta_2,\,
l_2 \zeta_3,\,
l_2 \zeta_2^2,\,
\right. \nonumber \\ && \left. \,\,\,\,
l_2^2 \zeta_2,\,
l_2^2 \zeta_2^2,\,
l_2^3 \zeta_2,\,
l_2^4 \zeta_2,\,
l_2 \zeta_2 \zeta_3,\,
a_4,\,
a_5,\,
\zeta_2 a_4,\,
s_6
\right\}
\end{eqnarray}
and three further constants ${\tilde \kappa}_1$, ${\tilde \kappa}_2$, ${\tilde \kappa}_3$ contribute.
It is worth mentioning that the coefficients of the intermediate expansions given in Eq.~(\ref{eq:expansion.at.s=infinity}) were complex numbers.
The real parts of these numbers also included the constants
\begin{eqnarray}
\label{eq:const-real-ori}
&& \left\{
l_2\,, l_2^2\,, l_2^3\,, l_2^2 \zeta_3,\,
\frac{1}{\pi^2},\, \frac{\zeta_3}{\pi^2}\,, \frac{\zeta_5}{\pi^2},\,
\frac{\zeta_3^2}{\pi^2},\, \frac{\zeta_3 \zeta_5}{\pi^2}\,,
\frac{l_2^4}{\pi^2},\, \frac{a_4}{\pi^2},\,  \frac{l_2^4 \zeta_3}{\pi^2},\, \frac{\zeta_3
a_4}{\pi^2},\,
\frac{{\tilde \kappa}_1}{\pi^2},\, \frac{{\tilde \kappa}_2}{\pi^2},\, \frac{{\tilde
\kappa}_3}{\pi^2}
\right\},
\end{eqnarray}
while the imaginary parts could be expressed as rational linear combinations of the following constants,
\begin{eqnarray}
\label{eq:const-imaginary-ori}
&& \left\{
\pi,\, \pi^3,\, \pi^5,\, \pi l_2,\, \pi^3 l_2,\, \pi^5 l_2,\, \pi l_2^2,\, \pi^3 l_2^2,\, \pi l_2^3,\, \pi^3 l_2^3,\, \pi l_2^4,\, \pi l_2^5,\,
\right. \nonumber \\ && \left. \,\,\,\,
\pi a_4,\, \pi l_2 a_4,\, \pi \zeta_3,\, \pi^3 \zeta_3,\, \pi l_2 \zeta_3,\, \pi l_2^2 \zeta_3
\right\}.
\end{eqnarray}

After combining the results in Eq.~(\ref{eq:F2fjl-s0}), the constants listed 
in Eq.~(\ref{eq:const-real-ori}) cancel out from the final result.
Furthermore, the imaginary contributions, expressed in terms of the 
constants in Eq.~(\ref{eq:const-imaginary-ori}), cancel exactly, leaving only 
real coefficients.

The different constants can be represented as iterated integrals,
\begin{eqnarray}
\HA_{b,\vec{a}}(1) = \int_0^1 dx f_b(x) \HA_{\vec{a}}(x)
\end{eqnarray}
of letters $f_c(x)$ in the alphabet $\mathfrak{A}$,
\begin{eqnarray}
\label{eq:hpl}
\mathfrak{A} = 
\Biggl\{f_0(x), f_1(x), f_{-1}(x) \Biggr\} \equiv
\Biggl\{\frac{1}{x}, \frac{1}{1-x}, \frac{1}{1+x}\Biggr\}.
\end{eqnarray}
Here 
\begin{eqnarray}
l_2     &=& \HA_{-1}(1), \\
a_4 &=&
-\frac{1}{24} \HA_{-1}(1)^4+\frac{1}{4} \HA_{-1}(1)^2 \HA_{0,1}(1)
+\frac{1}{10} \HA_{0,1}(1)^2
+\frac{1}{2} \HA_{0,0,-1,1}(1), \\
a_5 &=&
\frac{1}{120} \HA_{-1}(1)^5
-\frac{1}{12} \HA_{-1}(1)^3 \HA_{0,1}(1)
-\frac{19}{80} \HA_{-1}(1) \HA_{0,1}(1)^2
+\frac{7}{32} \HA_{-1}(1)^2 \HA_{0,0,1}(1)
\nonumber\\ &&
-\frac{3}{16} \HA_{0,1}(1) \HA_{0,0,1}(1)
+\frac{153}{128} \HA_{0,0,0,0,1}(1)
+\frac{1}{4} \HA_{0,0,-1,-1,1}(1).
\end{eqnarray}

We first note the occurrence of the constant
\begin{eqnarray}
\label{eq:C0}
s_6 = \zeta_6 - \HA_{0,0,0,0,-1,1}(1),~~~\zeta_6 = \HA_{0,0,0,0,0,1}(1)
\end{eqnarray}
in a final physical expression, with $\HA_{\vec{a}}(x)$ harmonic polylogarithms 
\cite{Remiddi:1999ew}.

Since in the color--planar case \cite{Ablinger:2018zwz} cyclotomic harmonic 
polylogarithms
\cite{Ablinger:2011te} of cyclotomy {\tt c = 6} contributed we searched for corresponding 
constants of this kind \cite{Ablinger:2011te,Broadhurst:1998rz}.
The constants ${\tilde \kappa}_1$ to ${\tilde \kappa}_3$ were found as the following iterative integrals
\begin{eqnarray}
\label{eq:C1}
{\tilde \kappa}_1 &=& \frac{3^5}{2^2} \HA_{\{3,0\},0,0}(1) = \pi^3 \sqrt{3} = 
53.704446563817351297, \\
{\tilde \kappa}_2 &=& \frac{3^5}{2^4} \HA_{\{3,0\},0,0}(1) \left[
\frac{3}{2} \HA_{\{3,0\},0}(1) - \HA_{\{6,0\},0}(1)\right]
= \pi^3 {\rm Cl}_2\left(\frac{\pi}{3}\right) = 31.469560262685681691,
\nonumber\\
\\
\label{eq:C3}
{\tilde \kappa}_3 &=& \frac{3^5}{2^2} \HA_{\{3,0\},0,0}(1) \HA_{\{\{4,-2,1\},0\},0,0}(1)
=
\pi^3 {\sf Im} {\rm \Li_3}\left[\frac{1}{2} e^{i\pi/3}\right] = 14.233090050526350538.
\end{eqnarray}
Terms of the kind  $\pi^3 \sqrt{3}$ are given by
\begin{eqnarray}
\label{eq:kap1}
\pi^3 \sqrt{3} = \frac{3^2}{2^3}\left[
\psi^{(2)}\left(\frac{2}{3}\right)  - \psi^{(2)} \left(\frac{1}{3}\right) \right],
\end{eqnarray}
with $\psi(x)$ the di-gamma function. Eq.~(\ref{eq:kap1})
can be expressed by the cyclotomic numbers
\begin{eqnarray}
\sum_{k=0}^\infty \frac{1}{(lk+m)^n} = \frac{1}{\Gamma(n)}\left(
- \frac{1}{l}\right)^n \psi^{n-1}\left(\frac{m}{l}\right),
\end{eqnarray}
related to the Hurwitz $\zeta$-function \cite{HURWITZ}.
The function ${\rm Cl}_2(\varphi)$ denotes the Clausen function 
\cite{CLAUSEN,LEWIN1,LEWIN2}
\begin{eqnarray}
{\rm Cl}_2(\varphi) = \sum_{k=1}^\infty \frac{\sin(k \varphi)}{k^2} = - \int_0^\varphi
\ln\left|2 \sin\left(\frac{x}{2}\right)\right| dx.
\end{eqnarray}
The letters forming the iterated integrals in  
Eqs.~(\ref{eq:C0},\ref{eq:C1}--\ref{eq:C3}) are finally given by those of $\mathfrak{A}$ 
and
\begin{eqnarray}
\label{eq:cyc}
f_{\{3,0\}}(x) &=& \frac{1}{1 + x + x^2},~~~
f_{\{6,0\}}(x) = \frac{1}{1 - x + x^2},\\
\label{eq:noncyc}
f_{\{4,-2,1\},0}(x) &=& \frac{1}{4 - 2x + x^2}.
\end{eqnarray}
and are either cyclotomic (\ref{eq:cyc}) or are numbers related to letters induced 
by quadratic forms, (\ref{eq:noncyc}), cf.~Ref.~\cite{Ablinger:2021fnc}. The letters~\eqref{eq:hpl} span the harmonic polylogarithms.

As examples we are showing the asymptotic expansions of the unrenormalized ${O}(\ep^0)$ 
contributions of the form factors $F_{V,1}$ and $F_{V,2}$
up to ${O}(1/s^5)$. Setting $N_c=3$, and defining ${\rm ls} = -\ln(-s)$, they are given by
\begin{eqnarray}
F_{V,1}^{s \rightarrow \infty} &=& 64 \Biggl\{
\Biggl[-\frac{1}{27}+\frac{49}{12960 s}-\frac{47}{810 s^2}-\frac{259}{1440 s^3}-\frac{491}{1944 s^4}+\frac{14189}{9720 s^5}\Biggr] {   \rm ls}^6
\nonumber\\ &&
+
\Biggl[\frac{119}{432}-\frac{407}{480 s}-\frac{16051}{12960 s^2}-\frac{15751}{3888 s^3}-\frac{945889}{77760 s^4}-\frac{2023951}{97200 s^5}\Biggr] {   \rm ls}^5
\nonumber\\ && 
+\Biggl[\left(-\frac{20}{81}+\frac{3673}{1296 s}+\frac{10195}{288 s^2}+\frac{197359}{648 s^3}+\frac{2660755}{1296 s^4}+\frac{7902395}{648 s^5}\right) \zeta_2
\nonumber\\ &&
-\frac{1681}{864 s}-\frac{571873}{10368 s^2}-\frac{58338557}{116640 s^3}-\frac{2353102207}{680400 s^4}-\frac{679446541129}{32659200 s^5}+\frac{3143}{1728}\Biggr] {   \rm ls}^4
\nonumber\\ &&
+\Biggl[\left(\frac{977}{648}-\frac{12475}{648 s}-\frac{118505}{432 s^2}-\frac{1264817}{486 s^3}-\frac{349389229}{19440 s^4}-\frac{52045981}{486 s^5}\right) \zeta_2
\nonumber\\ &&
+\Biggl(
\frac{110531688486577}{457228800}
-\frac{9532379}{162} \zeta_3
\Biggr)\frac{1}{s^5}
+\Biggl(
\frac{23621245300801}{571536000} 
-\frac{3261179}{324} \zeta_3
\Biggr)\frac{1}{s^4}
\nonumber\\ &&
+\Biggl(
\frac{21800859401}{3499200}
-\frac{41224}{27} \zeta_3
\Biggr)\frac{1}{s^3}
+\Biggl(\frac{5819821}{7776}       
-\frac{120311}{648} \zeta_3
\Biggr)\frac{1}{s^2}
\nonumber\\ &&
+\Biggl(
\frac{107167}{1296} 
- \frac{475}{36} \zeta_3
\Biggr)\frac{1}{s}
-\frac{254}{81} \zeta_3
+\frac{7579}{2592}\Biggr] {   \rm ls}^3
\nonumber\\ &&
+\Biggl[
\Biggl(-\frac{97}{54}-\frac{44093}{2160 s}-\frac{516317}{2160 s^2}-\frac{143461}{72 s^3}-\frac{811891}{60 s^4}-\frac{89472121}{1080 s^5}\Biggr) \zeta_2^2
\nonumber\\ &&  
+\frac{527}{54} \zeta_3
+\Biggl(
\frac{2140}{27} \zeta_3
+ \frac{107609}{324}
\Biggr)\frac{1}{s}
+\Biggl(
\frac{249451}{144} \zeta_3
+\frac{40985165}{31104}
\Biggr)\frac{1}{s^2}
+\Biggl(
\frac{9939443}{648} \zeta_3
\nonumber\\ &&
-\frac{6965769779}{11664000}
\Biggr)\frac{1}{s^3}
+\Biggl(
\frac{225487789}{2160} \zeta_3
-\frac{1207505059525747}{30005640000}
\Biggr)\frac{1}{s^4}
+\Biggl(
\frac{805169507}{1296} \zeta_3
\nonumber\\ &&
-\frac{1045382913180275311}{2880541440000}
\Biggr)\frac{1}{s^5}
-\frac{605705}{7776}
+\Biggl(
-\frac{2}{3} l_2
-\frac{13537}{216 s}
+\Biggl(\frac{29}{9} l_2
\nonumber\\ &&
-\frac{111209}{144}\Biggr)\frac{1}{s^2}
+\Biggl(
\frac{158}{9} l_2
-\frac{138093007}{38880}
\Biggr)\frac{1}{s^3}
+\Biggl(
\frac{2482}{27} l_2
-\frac{5152806637}{453600}
\Biggr)\frac{1}{s^4}
\nonumber\\ &&
+\Biggl(
\frac{7028}{15} l_2
-\frac{18453162643}{1088640}
\Biggr)\frac{1}{s^5}
-\frac{3979}{288}
\Biggr) \zeta_2
\Biggr] {   \rm ls}^2
\nonumber\\ &&
+\Biggl[
\frac{2}{9} l_2^4
+\Biggl(
\frac{8573}{1080}
+\frac{29899}{120 s}
+\frac{390703}{144 s^2}
+\frac{75849491}{3240 s^3}
+\frac{549568937}{3600 s^4}
+\frac{28296209321}{32400 s^5}\Biggr) \zeta_2^2
\nonumber\\ && 
+\frac{16}{3} a_4
+\zeta_2 \Biggl(
\frac{8}{3} l_2^2
-\frac{22}{3} l_2
+\Biggl(
\frac{631208}{405} l_2^2
-\frac{1303548121}{25200} l_2
\nonumber\\ && 
-\frac{748013}{54} \zeta_3
-\frac{2039036264236301}{1143072000}\Biggr)\frac{1}{s^5}
+\Biggl(\frac{30848}{81} l_2^2
-\frac{326120357}{37800} l_2
-\frac{15317}{6} \zeta_3
\nonumber\\ &&
-\frac{45991144470073}{142884000}\Biggr)\frac{1}{s^4}
+\Biggl(
\frac{2432}{27} l_2^2
-\frac{688471}{540} l_2
-\frac{34783}{108} \zeta_3
-\frac{30583655261}{583200}\Biggr)\frac{1}{s^3}
\nonumber\\ &&
+\Biggl(
\frac{560}{27} l_2^2
-\frac{8153}{54} l_2
+\frac{121}{432} \zeta_3
-\frac{18626471}{2592}\Biggr)\frac{1}{s^2}
+\Biggl(
-\frac{20}{27} l_2^2
-\frac{83}{9} l_2
\nonumber\\ &&
-\frac{905}{216} \zeta_3
-\frac{307519}{432}\Biggr)\frac{1}{s}
-\frac{37}{27} \zeta_3
-\frac{27295}{288}\Biggr)
+\frac{5113}{36} \zeta_3
+\Biggl(
\frac{76154}{1215} l_2^4
+\frac{609232}{405} a_4
\nonumber\\ &&
-\frac{28539584419}{21600} \zeta_3
-\frac{17712283}{108} {\zeta_5}
+\frac{50172038578234115621}{18517766400000}\Biggr)
\frac{1}{s^5}
+\Biggl(
\frac{3053}{243} l_2^4
\nonumber\\ &&
+\frac{24424}{81} a_4
-\frac{127902022793}{680400} \zeta_3
-\frac{3435319}{108} {\zeta_5}
+\frac{7652263475838252013}{16803158400000}\Biggr)\frac{1}{s^4}
\nonumber\\ &&
+\Biggl(
\frac{197}{81} l_2^4
+\frac{1576}{27} a_4
-\frac{6478301}{324} \zeta_3
-\frac{1121707}{216} {\zeta_5}
+\frac{1975638199433}{29160000}\Biggr)\frac{1}{s^3}
\nonumber\\ &&
+\Biggl(
\frac{67}{162} l_2^4
+\frac{268}{27} a_4
-\frac{334127}{432} \zeta_3
-\frac{520027}{864} {\zeta_5}
+\frac{47577215}{5832}\Biggr)
\frac{1}{s^2}
\nonumber\\ &&
+\Biggl(
\frac{10}{81} l_2^4
+\frac{80}{27} a_4
-\frac{755}{27} \zeta_3
-\frac{9665}{432} {\zeta_5}
+\frac{2940227}{3888}\Biggr)
\frac{1}{s}
+\frac{8}{9} {\zeta_5}
-\frac{4957}{9}\Biggr] {   \rm ls} 
\nonumber\\ &&
-\frac{16}{405} l_2^5
-\frac{7705}{1944} l_2^4
+\Biggl(
\frac{80947}{11340}+\frac{5891705}{81648 s}+\frac{33439039}{34020 s^2}+\frac{36748981}{4536 s^3}+\frac{257579593}{4860 s^4} 
\nonumber\\ && 
+\frac{12552412975}{40824 s^5}\Biggr) \zeta_2^3
+\Biggl(
-\frac{5}{9} l_2^2
+\frac{32}{135} l_2
+\Biggl(
\frac{667}{108} l_2^2
-\frac{244}{135} l_2
+\frac{6839519}{6480}\Biggr)\frac{1}{s}
\nonumber\\ &&
+\Biggl(
\frac{6185}{72} l_2^2
+\frac{874}{27} l_2
+\frac{168588197}{12960}\Biggr)\frac{1}{s^2}
+\Biggl(
\frac{14519}{18} l_2^2
+\frac{7568}{45} l_2
\nonumber\\ && 
+\frac{37711528277}{388800}\Biggr)\frac{1}{s^3}
+\Biggl(
\frac{154076}{27} l_2^2
+\frac{116132}{135} l_2
+\frac{24875134912819}{40824000}\Biggr)\frac{1}{s^4}
\nonumber\\ && 
+\Biggl(
\frac{104360}{3} l_2^2
+\frac{2853992}{675} l_2
+\frac{564913601711447}{163296000}\Biggr)\frac{1}{s^5}
+\frac{143363}{1440} \Biggr) \zeta_2^2
+\frac{199}{288} \zeta_3^2
\nonumber\\ && 
-\frac{7705}{81} a_4
+\frac{128}{27} a_5
+\frac{144233}{432} \zeta_3
+\zeta_2 \Biggl(
\frac{5}{54} l_2^4
+\frac{32}{81} l_2^3
-\frac{6235}{162} l_2^2
\nonumber\\ && 
+\Biggl(
\frac{35}{18} \zeta_3
+\frac{6965}{54}\Biggr) l_2
+\frac{20}{9} a_4
-\frac{19067}{432} \zeta_3
+\Biggl(
-\frac{667}{648} l_2^4
-\frac{64}{81} l_2^3
\nonumber\\ && 
-\frac{358}{9} l_2^2
+\Biggl(282
-\frac{1463}{72} \zeta_3
\Biggr) l_2
-\frac{667}{27} a_4
+\frac{148987}{648} \zeta_3
-\frac{90995}{54}\Biggr)\frac{1}{s}
\nonumber\\ &&
+\Biggl(
-\frac{5801}{432} l_2^4
-\frac{160}{81} l_2^3
+\frac{340}{9} l_2^2
+\Biggl(
\frac{176129}{162}
-\frac{41503}{144} \zeta_3
\Biggr) l_2
\nonumber\\ && 
-\frac{5801}{18} a_4
+\frac{152855}{54} \zeta_3
-\frac{97581929}{7776}\Biggr)\frac{1}{s^2}
+\Biggl(
-\frac{40037}{324} l_2^4
-\frac{256}{9} l_2^3
\nonumber\\ &&
+\frac{1419908}{1215} l_2^2
+\Biggl(\frac{32044222}{6075}
-\frac{99673}{36} \zeta_3
\Biggr) l_2
-\frac{80074}{27} a_4
\nonumber\\ &&
+\frac{32068373}{1296} \zeta_3
-\frac{108629217289}{1458000}\Biggr)\frac{1}{s^3}
+\Biggl(
-\frac{7822}{9} l_2^4
-\frac{17408}{81} l_2^3
+\frac{395239748}{42525} l_2^2
\nonumber\\ &&
+\Biggl(\frac{42591525673}{1488375}
-\frac{177982}{9} \zeta_3
\Biggr) l_2
-\frac{62576}{3} a_4
+\frac{3239859149}{19440} \zeta_3
\nonumber\\ && 
-\frac{129811254513023}{333396000}\Biggr) \frac{1}{s^4}
+\Biggl(
-\frac{428180}{81} l_2^4
-\frac{108416}{81} l_2^3
+\frac{490726262}{8505} l_2^2
\nonumber\\ && 
+\Biggl(\frac{14829601479241}{95256000}
-\frac{1090390}{9} \zeta_3
\Biggr) l_2
-\frac{3425440 a_4}{27}
+\frac{38342426077}{38880} \zeta_3
\nonumber\\ && 
-\frac{683253287727603317}{360067680000}\Biggr)\frac{1}{s^5}
+\frac{927827}{2592}\Biggr)
+\Biggl(
\frac{99298}{675} l_2^5
-\frac{105124618397}{8164800} l_2^4
\nonumber\\ && 
-\frac{4267943}{216} \zeta_3^2
-\frac{21441655613}{94478400} {{\tilde \kappa}_1}
-\frac{3247522}{10935} {{\tilde \kappa}_2}
-\frac{82880}{81} {{\tilde \kappa}_3}
-\frac{105124618397}{340200} a_4
\nonumber\\ && 
-\frac{794384}{45} a_5
-\frac{1302560}{81} {s_6}
+\frac{1708229617067371}{1143072000} \zeta_3
-\frac{2195310543401}{233280} {\zeta_5}
\nonumber\\ &&
-\frac{781202250523853512716013}{571643448768000000}\Biggr)\frac{1}{s^5}
+\Biggl(
\frac{9709}{405} l_2^5
-\frac{9482853731}{4082400} l_2^4
-\frac{435829}{144} \zeta_3^2
\nonumber\\ && 
-\frac{1501427}{34992} {{\tilde \kappa}_1}
-\frac{105527}{2187} {{\tilde \kappa}_2}
-\frac{13280}{81} {{\tilde \kappa}_3}
-\frac{9482853731}{170100}  a_4
-\frac{77672}{27} a_5
-\frac{229760}{81} {s_6}
\nonumber\\ &&
+\frac{52622091553511}{190512000} \zeta_3
-\frac{192516053233}{116640} {\zeta_5}
-\frac{113809853554807624541}{705732652800000}
\Biggr)\frac{1}{s^4}
\nonumber\\ &&
+\Biggl(
\frac{49}{15} l_2^5
-\frac{22727233}{58320} l_2^4
-\frac{266927}{864} \zeta_3^2
-\frac{401129}{52488} {{\tilde \kappa}_1}
-\frac{4820}{729} {{\tilde \kappa}_2}
-\frac{22727233}{2430} a_4
\nonumber\\ &&
-\frac{1760}{81} {{\tilde \kappa}_3}
-392 a_5 
-\frac{36320}{81} {s_6}
+\frac{2637615317}{58320} \zeta_3
-\frac{2012950841}{7776} {\zeta_5}
\nonumber\\ &&
-\frac{32336724212507}{3936600000}\Biggr)\frac{1}{s^3}
+\Biggl(
\frac{41}{162} l_2^5
-\frac{38467}{648} l_2^4
+\frac{3473}{1152} \zeta_3^2
-\frac{9157}{8748} {{\tilde \kappa}_1}
\nonumber\\ &&
-\frac{143}{243} {{\tilde \kappa}_2}
-\frac{16}{9} {{\tilde \kappa}_3}
-\frac{38467}{27} a_4
-\frac{820}{27} a_5
-\frac{512}{9} {s_6}
+\frac{5062873}{864} \zeta_3
-\frac{10666681}{324} {\zeta_5}
\nonumber\\ &&
+\frac{1161351943}{559872}\Biggr)\frac{1}{s^2}
+\Biggl(
\frac{32}{405} l_2^5
-\frac{893}{108} l_2^4
-\frac{683}{64} \zeta_3^2
-\frac{13}{243} {{\tilde \kappa}_1}
+\frac{5}{729} {{\tilde \kappa}_2}
\nonumber\\ &&
-\frac{1786}{9} a_4
-\frac{256}{27} a_5
-\frac{80}{27} {s_6}
+\frac{693095}{1296} \zeta_3
-\frac{9370877}{3888} {\zeta_5}
+\frac{37778}{81}\Biggr)\frac{1}{s}
\nonumber\\ &&
-\frac{318019}{864} {\zeta_5}
+\frac{788081}{7776} + {O}({\rm ls}^6/s^6)\Biggr\},
\end{eqnarray}

\begin{eqnarray}
F_{V,2}^{s \rightarrow \infty} &=& 64 \Biggl\{
\Biggl[-\frac{43}{2430 s^2}-\frac{1973}{19440 s^3}-\frac{521}{810 s^4}-\frac{31661}{9720 s^5}\Biggr] { \rm ls}^6
\nonumber\\ &&
+\Biggl[\frac{16}{27 s}+\frac{529}{360 s^2}+\frac{36737}{6480 s^3}+\frac{335527}{19440 s^4}+\frac{234701}{4860 s^5}\Biggr] { \rm ls}^5
\nonumber\\ &&
+\Biggl[\Biggl(\frac{11828169859}{583200}
-\frac{1728853}{162} \zeta_2
\Biggr)\frac{1}{s^5}
+\Biggl(\frac{389591423}{116640}
-\frac{31255}{18} \zeta_2
\Biggr)\frac{1}{s^4}
\nonumber\\ &&
+\Biggl(\frac{1231505}{2592}
-\frac{77591}{324} \zeta_2
\Biggr)\frac{1}{s^3}
+\Biggl(
\frac{33059}{648}
-\frac{1645}{72} \zeta_2
\Biggr)\frac{1}{s^2}
-\frac{205}{216 s}\Biggr] 
{ \rm ls}^4
\nonumber\\ &&
+\Biggl[\Biggl(
-\frac{206}{81} \zeta_2
-\frac{2683}{72}\Biggr)\frac{1}{s}
+\Biggl(149 \zeta_2
+\frac{8642}{81} \zeta_3
-\frac{51019}{108}\Biggr)\frac{1}{s^2}
+\Biggl(
\frac{23711}{12} \zeta_2
\nonumber\\ &&
+\frac{10225}{9} \zeta_3
-\frac{707251}{144}\Biggr)\frac{1}{s^3}
+\Biggl(
\frac{7290761}{486} \zeta_2
+\frac{223301}{27} \zeta_3
-\frac{10250948137}{291600}\Biggr)\frac{1}{s^4}
\nonumber\\ &&
+\Biggl(
\frac{38050922}{405} \zeta_2
+\frac{4109758}{81} \zeta_3
-\frac{817950539009}{3827250}\Biggr)\frac{1}{s^5}\Biggr] { \rm ls}^3
\nonumber\\ &&
+\Biggl[
\Biggl(
\frac{6424459}{90} \zeta_2^2
+\Biggl(
\frac{4865676071}{97200}
-\frac{6356}{27} l_2
\Biggr) \zeta_2
-\frac{437039791}{810} \zeta_3
\nonumber\\ &&
+\frac{969676241097559}{4286520000}\Biggr)\frac{1}{s^5}
+\Biggl(
\frac{166663}{15} \zeta_2^2
+\Biggl(
\frac{55830679}{3240}-40 l_2\Biggr) \zeta_2
-\frac{13961615}{162} \zeta_3
\nonumber\\ &&
+\frac{27772594783}{2187000}\Biggr)\frac{1}{s^4}
+\Biggl(
\frac{81133}{54} \zeta_2^2
+\Biggl(\frac{958817}{216}-6 l_2\Biggr) \zeta_2
-\frac{617723}{54} \zeta_3
-\frac{11218685}{2592}\Biggr)\frac{1}{s^3}
\nonumber\\ &&
+\Biggl(
\frac{15853}{108} \zeta_2^2
+\Biggl(\frac{22747}{27}
-\frac{4}{9} l_2
\Biggr) \zeta_2
-\frac{25177}{27} \zeta_3
-\frac{639259}{324}\Biggr)\frac{1}{s^2}
\nonumber\\ &&
+\Biggl(
\frac{251}{36} \zeta_2
+\frac{416}{27} \zeta_3
-\frac{1477}{9}\Biggr)\frac{1}{s}\Biggr] { \rm ls}^2
\nonumber\\ &&
+\Biggl[\Biggl(
-\frac{4}{81} l_2^4
+\frac{7157}{270} \zeta_2^2
-\frac{32}{27} a_4
+\left(\frac{8}{27} l_2^2
+\frac{8}{9} l_2
+\frac{51845}{216}\right) \zeta_2
\nonumber\\ &&
+\frac{2512}{27} \zeta_3
-\frac{205843}{648}\Biggr)\frac{1}{s}
+\Biggl(
-\frac{2}{81} l_2^4
-\frac{9492}{5} \zeta_2^2
-\frac{16}{27} a_4
+\frac{2779}{27} \zeta_3
\nonumber\\ &&
+\zeta_2 \Biggl(
\frac{16}{27} l_2^2
+\frac{256}{3} l_2
+\frac{1811}{54} \zeta_3
+\frac{56231}{12}\Biggr)
+\frac{51533}{108} \zeta_5
-\frac{453752}{81}\Biggr)\frac{1}{s^2}
\nonumber\\ &&
+\Biggl(
-\frac{7}{81} l_2^4
-\frac{5316391}{270} \zeta_2^2
-\frac{56}{27} a_4
+\frac{271039}{18} \zeta_3
+\zeta_2 
\Biggl(
\frac{176}{27} l_2^2
+\frac{25358}{27} l_2
\nonumber\\ &&
+\frac{11342}{27} \zeta_3
+\frac{8905771}{216}\Biggr)
+\frac{126890}{27} \zeta_5
-\frac{463049833}{7776}\Biggr)\frac{1}{s^3}
+\Biggl(
-\frac{164}{81} l_2^4
-\frac{55476766}{405} \zeta_2^2
\nonumber\\ &&
-\frac{1312}{27} a_4
+\frac{154072027}{972} \zeta_3
+\zeta_2 \Biggl(
\frac{1408}{27} l_2^2
+\frac{2865638}{405} l_2
+\frac{172445}{54} \zeta_3
\nonumber\\ &&
+\frac{19895774023}{72900}\Biggr)
+\frac{3284117}{108} \zeta_5
-\frac{3531518369599}{8201250}\Biggr)\frac{1}{s^4}
+\Biggl(
-\frac{158}{9} l_2^4
-\frac{2182382911}{2700} \zeta_2^2
\nonumber\\ &&
-\frac{1264}{3} a_4
+\frac{28314666569}{24300} \zeta_3
+\zeta_2 \Biggl(
\frac{9200}{27} l_2^2
+\frac{3304387}{75} l_2
+\frac{938033}{54} \zeta_3
\nonumber\\ &&
+\frac{32111077988851}{20412000}\Biggr)
+\frac{17634373}{108} \zeta_5
-\frac{1588111136598363089}{600112800000}\Biggr)\frac{1}{s^5}\Biggr] { \rm ls}
\nonumber\\ &&
+\Biggl(
-\frac{16}{405} l_2^5
+\frac{404}{81} l_2^4
+\Biggl(
\frac{122}{135} l_2
-\frac{124927}{360}\Biggr) \zeta_2^2
+\frac{3232}{27} a_4
+\frac{128}{27} a_5
\nonumber\\ &&
+\zeta_2 \Biggl(
\frac{32}{81} l_2^3
+\frac{992}{27} l_2^2
-\frac{1868}{9} l_2
-\frac{2305}{54} \zeta_3
+\frac{3495}{8}\Biggr)
+\frac{35819}{216} \zeta_3
+\frac{577}{9} \zeta_5
\nonumber\\ &&
-\frac{674143}{648}\Biggr)\frac{1}{s}
+\Biggl(
-\frac{166}{405} l_2^5
+\frac{3152}{81} l_2^4
-\frac{30180217}{51030} \zeta_2^3
+\Biggl(
-\frac{1003}{27} l_2^2
-\frac{628}{135} l_2
\nonumber\\ &&
-\frac{3800123}{405}\Biggr) \zeta_2^2
-\frac{21515}{432} \zeta_3^2
+\frac{260}{243} {\tilde \kappa}_1
+\frac{232}{729} {\tilde \kappa}_2
+\frac{32}{27} {\tilde \kappa}_3
+\frac{25216}{27} a_4
+\frac{1328}{27} a_5
\nonumber\\ &&
+\frac{1024}{27} {s_6}
-\frac{195793}{162} \zeta_3
+\zeta_2 \Biggl(
\frac{907}{162} l_2^4
+\frac{320}{81} l_2^3
-\frac{856}{27} l_2^2
+\Biggl(
\frac{2191}{18} \zeta_3
\nonumber\\ &&
-\frac{2072}{3}\Biggr) l_2
+\frac{3628}{27} a_4
-\frac{625295}{324} \zeta_3
+\frac{3021521}{324}\Biggr)
+\frac{40430395}{1944} \zeta_5
-\frac{1480295}{324}\Biggr)\frac{1}{s^2}
\nonumber\\ &&
+ \Biggl(-\frac{169}{45} l_2^5
+\frac{5201}{18} l_2^4
-\frac{321994651}{51030} \zeta_2^3
+\Biggl(
-\frac{16426}{27} l_2^2
-\frac{692}{15} l_2
\nonumber\\ &&
-\frac{259746577}{3240}\Biggr) \zeta_2^2
+\frac{97145}{648} \zeta_3^2
+\frac{15031}{2187} {\tilde \kappa}_1
+\frac{392}{81} {\tilde \kappa}_2
+\frac{4096}{243} {\tilde \kappa}_3
+\frac{20804}{3} a_4
\nonumber\\ &&
+\frac{1352}{3} a_5
+\frac{76672}{243} {s_6}
-\frac{1331983}{54} \zeta_3
+\zeta_2 \Biggl(
\frac{22591}{243} l_2^4
+\frac{320}{9} l_2^3
\nonumber\\ &&
-\frac{28792}{27} l_2^2
+\Biggl(
\frac{6307}{3} \zeta_3
-\frac{293876}{81}\Biggr) l_2
+\frac{180728}{81} a_4
-\frac{2156093}{108} \zeta_3
\nonumber\\ &&
+\frac{81713761}{1296}\Biggr)
+\frac{43715215}{216} \zeta_5
-\frac{189711101}{46656}\Biggr)\frac{1}{s^3}
+\Biggl(
-\frac{2092}{81} l_2^5
+\frac{13572359}{7290} l_2^4
\nonumber\\ &&
-\frac{383089676}{8505} \zeta_2^3
+\Biggl(
-\frac{43879}{9} l_2^2
-\frac{8200}{27} l_2
-\frac{51806903917}{97200}\Biggr) \zeta_2^2
+\frac{1076027}{432} \zeta_3^2
\nonumber\\ &&
+\frac{562607}{13122} {\tilde \kappa}_1
+\frac{9460}{243} {\tilde \kappa}_2
+\frac{11200}{81} {\tilde \kappa}_3
+\frac{54289436}{1215} a_4
+\frac{83680}{27} a_5
+\frac{175360}{81} {s_6}
\nonumber\\ &&
-\frac{5509262849}{29160} \zeta_3
+\zeta_2 \Biggl(
\frac{120437}{162} l_2^4
+\frac{19840}{81} l_2^3
-\frac{10613552}{1215} l_2^2
+\Biggl(
\frac{102011}{6} \zeta_3
\nonumber\\ &&
-\frac{129789793}{6075}\Biggr) l_2
+\frac{481748}{27} a_4
-\frac{46607033}{324} \zeta_3
+\frac{380815246837}{1093500}\Biggr)
\nonumber\\ &&
+\frac{100578085}{72} \zeta_5
+\frac{64087983862793}{656100000}\Biggr)\frac{1}{s^4}
+\Biggl(
-\frac{20846}{135} l_2^5
+\frac{793387189}{72900} l_2^4
\nonumber\\ &&
-\frac{2309825303}{8505} \zeta_2^3
+\Biggl(
-\frac{830825}{27} l_2^2
-\frac{235204}{135} l_2
-\frac{4553200031557}{1458000}\Biggr) \zeta_2^2
\nonumber\\ &&
+\frac{7798951}{432} \zeta_3^2
+\frac{1319033}{6561} {\tilde \kappa}_1
+\frac{184720}{729} {\tilde \kappa}_2
+\frac{24320}{27} {\tilde \kappa}_3
+\frac{1586774378}{6075} a_4
\nonumber\\ &&
+\frac{166768}{9} a_5
+\frac{116480}{9} {s_6}
-\frac{107122694929}{94500} \zeta_3
+\zeta_2 \Biggl(
\frac{757865}{162} l_2^4
\nonumber\\ &&
+\frac{118720}{81} l_2^3
-\frac{334037996}{6075} l_2^2
+\Biggl(
\frac{1938965}{18} \zeta_3
-\frac{8773381222}{70875}\Biggr) l_2
\nonumber\\ &&
+\frac{3031460}{27} a_4
-\frac{714704789}{810} \zeta_3
+\frac{3003044947049837}{1714608000}\Biggr)
+\frac{40345127863}{4860} \zeta_5
\nonumber\\ &&
+\frac{79286374540590832453}{75614212800000}\Biggr)\frac{1}{s^5} + {O}({\rm 
ls}^6/s^6)\Biggr\}.
\end{eqnarray}


\noindent
The corresponding expressions for $s \rightarrow +\infty$ are obtained by setting
\begin{eqnarray}
\label{eq:repl}
\ln\left(-{s}\right) \rightarrow  \ln\left(s\right) + i \pi
\end{eqnarray}
in the expressions for $s \rightarrow -\infty$.

We have tried to guess the polynomial parts in Eq.~(\ref{eq:x0}) for all  
form factor expansions up to ${O}(1/s^{1400})$ as closed functions of $s$.
This has been possible for the most simple contributions. The respective terms yield the representation
of the form factor within the convergence radius of the series in closed form. Due to the presence of 
a threshold at $s=4$ one expects
square-root valued functions. First one obtains infinite sum representations, containing central binomials
$\binom{2n}{n}$. These can be handled by algorithms of  Ref.~\cite{Ablinger:2014bra} 
and Ref.~\cite{Blumlein:2023vwi}.
Some of these sum-types have also been considered in the earlier literature 
\cite{Davydychev:2003mv,Weinzierl:2004bn}.
A simple example is 
\begin{eqnarray}
\label{eq:exbinom}
\lefteqn{\sum_{n=0}^\infty \frac{(-27 + 101 n - 94 n^2 - 371 n^3 + 754 n^4 - 644 n^5 + 
216 n^6)}{
12 (2 n - 3) (2 n - 1)^2 (1 + 2 n)} \binom{2n}{n} \left(-\frac{1}{s}\right)^n  }
\nonumber\\ && \hspace*{4cm} =
\frac{5 s^4-36 s^3+146 s^2-68 s+256}{6 (s-4)^{5/2} s^{3/2}}+\frac{(s-16)
   {\rm arcsin}\left(\frac{2}{\sqrt{s}}\right)}{24 \sqrt{s}}.
\nonumber\\
\end{eqnarray} 
containing the threshold $\sqrt{s - 4}$ and the pseudo-threshold factor $(s - 16)$.
In the high energy result, logarithms $\ln(\sqrt{4-s})$, 
cf.~Section~\ref{Sec:NumericalResults}, have been expanded already. This expression signals that 
the expansion around $s \rightarrow \infty$ will diverge at $s = 4$, which is confirmed
by applying Cauchy's criterion on the summand in the l.h.s. of Eq.~(\ref{eq:exbinom}).

From about 1400 coefficients one can obtain both recurrences and differential equations 
describing 
the respective color-$\zeta$ values and powers in 
$\ln\left(-{1}/{s}\right)$. For the recurrences only a few are
first order factorizing and one finds closed form solutions like (\ref{eq:exbinom}).

Here it is convenient to use the variable $x$ instead of $s$. In the sum-expressions
also cyclotomic harmonic sums \cite{Ablinger:2011te} 
\begin{eqnarray}
S_{\{a,b,c\}}({n}) =
\sum_{k=1}^N \frac{1}{(a k + b)^c},~~~a,b \in \mathbb{Z} \backslash \{0\},~~~c \in
\mathbb{N}
\end{eqnarray}
occur. By using the command {\tt ComputeGeneratingFunction} of {\tt HarmonicSums} 
\cite{Ablinger:2010kw,Ablinger:2013hcp,Ablinger:2014bra,Ablinger:2015tua}, 
the result is obtained by finding the corresponding differential equation of the 
problem. To keep the degrees of the differential equations low, larger polynomials in $n$
should be partial fractioned.
The differential equations are then solved in iterating integrals over the letters of the 
corresponding alphabet, which contains 
Kummer-Poincar\'e \cite{KUMMER1,KUMMER2,KUMMER3,
POINCARE1,LAPPO,CHEN,GONCHAROV,Moch:2001zr,Ablinger:2013cf}
\begin{eqnarray}
\frac{1}{x - c},~~~c \in \mathbb{C},
\end{eqnarray}
and square-root valued letters
\begin{eqnarray}
\sqrt{x} \sqrt{a + x},~~~{\rm etc.}
\end{eqnarray}
We consider one example of the high multiple zeta value 
$C_F^3 \zeta_2 l_2^2 {\rm ls}$. 
The corresponding recurrence is expected to be still relatively simple. One obtains
\begin{eqnarray}
\hspace*{-1cm}  
\sum_{n=0}^{\infty } \left(
        -\frac{1}{s}\right)^n && \hspace*{-10mm} \Biggl[
        \frac{\displaystyle (-1)^n \big(
                27-135 n+266 n^2-266 n^3+176 n^4-56 n^5\big) \binom{2 n}{n}}
{6 (2 n-3) (2 n-1)^2} 
\nonumber\\ && \hspace*{-1cm}
        +\frac{\displaystyle (-1)^n n \binom{2 n}{n} S_{\{2,-1,1\}}({n})}{3 (2 n-1)}
\Biggr] 
= -\frac{P_1}{6 (1-x)^3 (1+x)^5}
+\frac{\displaystyle 2 x \ln \left(
        \frac{1+x}{1-x}\right)}{3 (1-x^2)},
\\
P_1&=& 9 - 2 x + 88 x^2 - 238 x^3 + 670 x^4 - 238 x^5 + 88 x^6 - 2 x^7 + 9 x^8.
\end{eqnarray} 
Here the  solution depends only on the harmonic polylogarithms $\HA_1(x)$ and 
$\HA_{-1}(x)$, after rewriting the 
Kummer-Poincar\'e integrals with a more complicated main variable accordingly. The 
iterated integrals with root-valued letters only occur in intermediate steps of the 
calculation.

Let us guess the corresponding difference equations and investigate their 
factorization, cf.~Ref.~\cite{Ablinger:2018cja}.
We consider the case of the $C_F^3$ term at ${O}(\ep^0)$ of the form factor $F_{V,1}(s)$. 
The available number of
coefficients allowed to obtain 28 recurrences, out of which 10 are first order factorizing,
as illustrated in Table~\ref{TAB1}.

\renewcommand*{\arraystretch}{1.2} 
\begin{table}[h]\centering 
{\footnotesize \begin{tabular}{|c|c|r|r|r|} \hline \hline 
\multicolumn{5}{|c|}{recurrence} \\ 
\hline \hline
\multicolumn{1}{|c|}{${\rm ls}^k$} & 
\multicolumn{1}{c|}{constant} & 
\multicolumn{1}{c|}{degree} & 
\multicolumn{1}{c|}{order} &
\multicolumn{1}{c|}{\# 1st ord.}
\\
\hline \hline
 ${\rm ls}^6$ &  1& 18& 5& 2 \\
 \hline
 ${\rm ls}^4$& $\zeta_2$& 18& 5& 2 \\ 
 \hline
 ${\rm ls}^3$& $\zeta_3$& 18& 5& 2\\ 
 \hline
 ${\rm ls}^2$ & $\zeta_2^2$& 18& 5& 2\\ 
 ${\rm ls}^2$& $\ln(2) \zeta_2$& 8& 2& 2\\ 
 \hline
 ls& $\Li_4(1/2)$& 8& 2& 2\\ 
 ls& $\zeta_5$& 18& 5& 2 \\
 ls& $\zeta_2 \zeta_3$& 18& 5& 2   \\
 ls& $\ln^2(2) \zeta_2$& 7& 2& 2  \\
 ls& $\ln(2) \zeta_2$& 126& 19& 9 \\ 
 ls& $\ln^4(2)$& 8& 2& 2 \\
 \hline
 1& $\ln^4(2)$& 153& 22& 9\\
 1& $\Li_4(1/2)$& 153& 22& 9\\ 
 1& $\ln^3(2) \zeta_2$& 5& 1& 1\\ 
 1& $\ln(2) \zeta_2^2$& 8& 2& 2\\ 
 1& $\ln^5(2)$& 8& 2& 2\\
 1& $\Li_5(1/2)$ & 8& 2& 2\\ 
 1& $\zeta_2^3$& 18& 5& 2\\ 
 1& $\ln(2) \zeta_2 \zeta_3$& 18& 5& 2\\ 
 1& $\ln^2(2) \zeta_2^2$& 18& 5& 2\\
 1& $\ln^4(2) \zeta_2$& 18& 5& 2\\
 1& $\zeta_3^2$& 18& 5& 2\\
 1& $\zeta_2 \Li_4(1/2)$& 18& 5& 2\\ 
 1& $s_6$& 3& 1& 1\\
 1& $\sqrt{3} \pi \zeta_2$& 186& 20& 9\\ 
 1& $\pi \zeta_2 {\sf Im}[\Li_2(e^{i \pi/3}]$ & 41& 11& 7\\
 1& $\ln^2(2) \zeta_2$& 60& 12& 5\\ 
 1& $\pi \zeta_2 {\sf Im}[\Li_3(1 + i \sqrt{3})/4]$& 1& 1& 1\\
 \hline
 \hline
\end{tabular}}
\caption[]{\sf The parameters of the guessed recurrences for $s \rightarrow \infty$ of 
the $C_F^3$ term of the form factor $F_{V,1}(s)$. \label{TAB1}}
\end{table}

For the solvable part around $s \rightarrow \infty$ of the $C_F^3$-term for the 
vector 
form factor $F_{V,1}(s)$ one obtains 
\begin{eqnarray}
\label{asyCF3V1}
\lefteqn{F_{V,1}^{s \rightarrow \infty, solv.} = 64 \Biggl\{\frac{C_F^3}{(1-x)^3 
(x+1)^5}} 
\nonumber\\ &&
\Biggl\{
\pi ^2 l_2 {\rm ls}^2 \Biggl[\frac{1}{8} \left(3 x^8+3 x^7-26 x^6-43 x^5-130 x^4-43 
x^3
-26 x^2+3 x+3\right)
\nonumber\\ &&
-\frac{1}{16} {\LL} (1-x)^2 (x+1)^2 \left(3 x^4+28 x^3+22 x^2+28 x+3\right)\Biggr]
\nonumber\\ &&
+{\rm ls} \Biggl[
\frac{1}{48} {\LL} l_2^4 (1-x)^2 (x+1)^2 \left(9 x^4+68 x^3+34 x^2+68 x+9\right)
\nonumber\\ && 
+ a_4
\Biggl(\frac{1}{2} {\LL} (1-x)^2 (x+1)^2 \left(9 x^4+68 x^3+34 x^2+68 x+9\right)
\nonumber\\ &&
-18 x^8-107 x^7+250 x^6-925 x^5+3136 x^4-925 x^3+250 x^2-107 x-18\Biggr)
\nonumber\\ &&
+\pi ^2 l_2^2 \Biggl(\frac{1}{6} \left(-9 x^8+2 x^7-88 x^6+238 x^5-670 x^4+238 x^3
-88 x^2+2 x-9\right)
\nonumber\\ &&
-\frac{2}{3} {\LL} (1-x)^2 x (x+1)^4\Biggr)
\nonumber\\ &&
-\frac{1}{24} {l_2^4} \left(18 x^8+107 x^7-250 x^6+925 x^5-3136 x^4+925 x^3-250 
x^2+107 x
+18\right)
\Biggr]
\nonumber\\ &&
+\frac{32}{3} \pi ^3 (1-x)^2 x^3 
{\sf Im}\left[\Li_3\left(\frac{1}{4} \left(1+i \sqrt{3}\right)\right)\right]
\nonumber\\ &&
+
a_5
\Biggl[\frac{3}{2} {\LL} (1-x)^2 (x+1)^2 \left(3 x^4+12 x^3-10 x^2+12 x+3\right)
\nonumber\\ &&
-16 x^8+233 x^7-1006 x^6+3791 x^5-8052 x^4+3791 x^3-1006 x^2+233 x-16\Biggr]
\nonumber\\ &&
+l_2^5 \Biggl[\frac{1}{120} \left(16 x^8-233 x^7+1006 x^6-3791 x^5+8052 x^4
-3791 x^3+1006 x^2-233 x+16\right)
\nonumber\\ &&
-\frac{1}{80} {\LL} (1-x)^2 (x+1)^2 \left(3 x^4+12 x^3-10 x^2+12 x+3\right)\Biggr]
\nonumber\\ &&
+\pi ^4 l_2 \Biggl[\frac{1}{180} \left(-4 x^8+227 x^7-724 x^6+1469 x^5-8688 x^4
+1469 x^3-724 x^2+227 x-4\right)
\nonumber\\ &&
-\frac{1}{120} {\LL} (1-x)^2 (x+1)^2 \left(x^2+16 x+1\right) \left(3 x^2+4 
x+3\right)\Biggr]
\nonumber\\ &&
+\frac{8}{3} {s_6} (1-x)^2 \left(3 x^4-54 x^3+70 x^2-54 x+3\right) x 
\nonumber\\ &&
-\frac{2}{9} \pi ^2 l_2^3
\left(x^8-14 x^7+64 x^6-242 x^5+510 x^4-242 x^3+64 x^2-14 x+1\right) 
\Biggr\}\Biggr\},
\end{eqnarray}


\noindent
serving as an illustration.
Here one has ${\rm ls} = - 2 \ln(1-x) + \ln(x)$, after transforming back to the variable $x$ 
by
which square-root valued iterated integrals are turned into harmonic polylogarithms.
The results for the other color factors and currents are similar. The ones for the vector
current are given in the ancillary files in complete form.
\section{Numerical results for threshold and pseudo-threshold} 
\label{Sec:NumericalResults}

\vspace*{1mm}
\noindent
As we explained in Section \ref{Sec:AnalyticCont}, the kinematic region $s \in [0, \infty[$ has also been mapped out by overlapping series 
expansions. Here we used the values listed in (\ref{eq:s0>0-w4}), (\ref{eq:s0>0}) or (\ref{eq:s0>0-full}) depending on the constant in Eq.~(\ref{eq:kappa-const}) under
consideration.

The values at $s \rightarrow +\infty$ are obtained by applying
the replacement Eq.~(\ref{eq:repl}) to Eq.~(\ref{eq:x0}).
The values $s = 4$ and $s = 16$ mark the two-particle threshold and the four-particle pseudo-threshold. For the  
two vector currents the most singular 
terms are\footnote{Eqs.~(\ref{eq:FV1-4p-prop}--\ref{eq:FV2-4p-prop}) should be interpreted with each color factor having a different
constant of proportionality.}
\begin{eqnarray}
\label{eq:THR1}
F_{V,1}^{s=4} &\propto& 
  \left[\frac{C_F^3}{(4-s)^{5/2}} 
+ \frac{C_A C_F^2}{(4-s)^{3/2}} 
+ \frac{C_F C_A^2}{\sqrt{4-s}} \right] \ln^3(\sqrt{4-s}),
\label{eq:FV1-4p-prop}
\\
F_{V,2}^{s=4} &\propto& 
  \left[\left(\frac{C_F^3}{(4-s)^{5/2}} 
+ \frac{C_A C_F^2}{(4-s)^{3/2}}\right) \ln(\sqrt{4-s}) 
+ \frac{C_F C_A^2}{\sqrt{4-s}} \right] \ln^2(\sqrt{4-s}).
\label{eq:FV2-4p-prop}
\end{eqnarray}
For all form factors, the highest logarithmic power at threshold is $\ln^3(\sqrt{4-s})$.
We set $z = \sign{4-s} \sqrt{4-s}$ and calculate the form factors for QCD ($N_c = 3$),
\begin{eqnarray}
\label{eq:V14}
\lefteqn{\hspace*{-0.5cm} F_{V,1}^{s = 4}(s) =} \nonumber\\ && \hspace*{-1cm} 
64 \Biggl\{\Biggl[\frac{67.020643276582255754}{z^5}
+\frac{196.14487321500817679}{z^3}
-\frac{170.34206114472744875}{z^2}
\nonumber\\ &&  \hspace*{-1cm}
+\frac{240.25908053757386093}{z}
+113.94722282821479673
-27.034756712385753673 z
\nonumber\\ &&  \hspace*{-1cm}
+9.9213893624530990807 z^2
-47.821614988072272099 z^3
\Biggr]\ln^3(z)
\nonumber\\ &&  \hspace*{-1cm}
+\Biggl[
-\frac{335.10321638291127877}{z^5}
-\frac{140.36770703771532258}{z^4}
-\frac{859.35492161827618220}{z^3}
\nonumber\\ &&   \hspace*{-1cm}
+\frac{527.84106500640866095}{z^2}
-\frac{708.69657264031675763}{z}
-230.81098800267870479
\nonumber\\ &&    \hspace*{-1cm}
+573.08459738167342666 z
-66.333837596732951750 z^2
\nonumber\\ &&     \hspace*{-1cm}
+246.09111998237592795 z^3
\Biggl] \ln^2(z)
+\Biggl[
\frac{1206.4946345102209621}{z^5}
\nonumber\\ &&      \hspace*{-1cm}
+\frac{935.78471358476881719}{z^4}
+\frac{3893.6868101680108198}{z^3}
-\frac{2704.8274033004820109}{z^2}
\nonumber\\ &&       \hspace*{-1cm}
+\frac{2196.4468782816843489}{z}
+419.33347768644876125
-2738.2988837485641725 z
\nonumber\\ &&   \hspace*{-1cm}
+182.05054208017950358 z^2
-771.12031243138006769 z^3
\Bigg] \ln(z)
\nonumber\\ &&  \hspace*{-1cm}
-\frac{1421.8495670486650925}{z^5}
-\frac{1681.6788169229116386}{z^4}
\nonumber\\ &&  \hspace*{-1cm}
-\frac{5201.6536874048176874}{z^3}
+\frac{3121.3715770767315534}{z^2}
-\frac{260.96293808394154879}{z}
\nonumber\\ &&  \hspace*{-1cm}
-3411.6235285341735782
+3671.2042388135175651 z
\nonumber\\ &&  \hspace*{-1cm}
-938.55790702760894830 z^2 
+772.18066397138656325 z^3
+ {O}(\ln^3(z) z^4)\Biggr\},
\\
\label{eq:V24}
\lefteqn{\hspace*{-0.5cm} F_{V,2}^{s = 4}(s) =} \nonumber\\ && \hspace*{-1cm} 
\nonumber\\ &&
64 \Biggl\{\Biggl[
-\frac{67.020643276582255754}{z^5}
+\frac{93.578471358476881719}{z^4}
-\frac{146.60765716752368446}{z^3}
\nonumber\\ &&  \hspace*{-1cm}
-\frac{70.183853518857661289}{z^2}
+\frac{8.0672996536626789333}{z}
+5.8486544599048051075
\nonumber\\ &&  \hspace*{-1cm}
+17.499107412808146921 z
+1.4621636149762012769 z^2
+4.4451765612620808251 z^3
\Biggr] \ln^3(z)
\nonumber\\ &&  \hspace*{-1cm}
+\Biggl[\frac{234.57225146803789514}{z^5}
-\frac{233.94617839619220430}{z^4}
+\frac{1267.0638348841515714}{z^3}
\nonumber\\ &&  \hspace*{-1cm}
-\frac{194.83330169557882014}{z^2}
+\frac{343.26860551225689033}{z}
+215.22439177610109378
\nonumber\\ &&  \hspace*{-1cm}
+86.629039642351514841 z
+51.361586667011192180 z^2
+13.310190423572710157 z^3
\Biggr] \ln^2(z)
\nonumber\\ &&  \hspace*{-1cm}
+\Biggl[
-\frac{770.86045321243629969}{z^5}
-\frac{81.752298987167572850}{z^4}
-\frac{4252.8858245363383188}{z^3}
\nonumber\\ &&  \hspace*{-1cm}
+\frac{812.50078143439908563}{z^2}
-\frac{668.55658351210289137}{z}
-671.91897951899863342
\nonumber\\ &&  \hspace*{-1cm}
-235.76866949907519536 z
-136.75818842948801771 z^2
+4.8315736060575201094 z^3
\Biggl] \ln(z)
\nonumber\\ &&  \hspace*{-1cm}
+\frac{651.05064160209897205}{z^5}
+\frac{688.55371120890422021}{z^4}
+\frac{6515.4518611427798796}{z^3}
\nonumber\\ &&  \hspace*{-1cm}
-\frac{1539.4430480039000202}{z^2}
+\frac{1443.8094142399721744}{z}
-185.89212672255244637
\nonumber\\ &&  \hspace*{-1cm}
+315.58943821132890552 z
+137.97759955857833395 z^2
-61.012626525594663345 z^3
\nonumber\\ &&  \hspace*{-1cm}
+ {O}(\ln^3(z) z^4)\Biggr\}.
\end{eqnarray}

For the representation around $s = 16$ we obtain for the unrenormalized vector form 
factors $F_{V,1}$ and $F_{V,2}$
\begin{eqnarray}
\label{eq:V16}
F_{V,1}^{s = 16}(s) &=& 64 \Bigg\{
~~(2325.4946308219634279-487.4315999866111024~i)
\nonumber\\ &&
+(22.480629736147760202+21.299424286130607709~i)~{\rr}
\nonumber\\ &&
-(0.8743653418526481576+1.8387376395950035924~i)~{\rr}^2
\nonumber\\ &&
+(0.12341341541186443744+0.15565964113939265584~i)~{\rr}^3
\nonumber\\ &&
-(0.013592443881038082748+0.011969209955373434833~i)~{\rr}^4
\nonumber\\ &&
-2.1838658955132801214947 \times 10^{-7}~i~{\rr}^{9/2} + {O}(\rr^5)\Biggr\},
\\
\label{eq:V26}
F_{V,2}^{s = 16}(s) &=& 64 \Biggl\{
-(212.54368142723435665-99.69718917811286925~i)
\nonumber\\ &&
+(8.529885254529849981-16.141024662980248233~i)~{\rr}
\nonumber\\ &&
+(0.1468751565593577835+1.4776029465283092534~i)~{\rr}^2
\nonumber\\ &&
-(0.072252433798853051314+0.117906955628220334307~i)~{\rr}^3
\nonumber\\ &&
+(0.0102592475669520651171+0.0088850413522024968023~i)~{\rr}^4
\nonumber\\ &&
+1.12144464904736006238919 \times 10^{-7}~i~{\rr}^{9/2}
+ {O}(\rr^5)\Biggr\}.
\end{eqnarray}
At $s = 4$ we calculated  4000 coefficients per logarithmic power with 1200 digits of
precision and, similarly, at $s = 16$, we obtained 4000 coefficients with 300 digits of precision.
Notice that in Eqs.~(\ref{eq:V16}) and~(\ref{eq:V26}) the lowest non-integer power is ${\rr}^{9/2}$ in both cases.
As we mentioned in Section~\ref{Sec:AnalyticCont}, the expansions around $s=16$ were obtained by matching them directly
with the asymptotic expansions at $s \rightarrow \infty$.

We have also performed a numerical comparison of our unrenormalized form factors with those of the numerical 
programme of  
Ref.~\cite{Fael:2022miw,Fael:2022rgm,KIT1,MATKIT} at selected values of 
\begin{eqnarray}
\label{eq:spoints}
s \in \left\{-\frac{1}{10}, -5, -10, -50, -200, 200, 16, 5, 3, 2, 1, 
\frac{1}{10} \right\}
\end{eqnarray}
spanning a wide range of the kinematic region.  We found relative agreement of better 
than $10^{-10}$.

\section{Conclusions} \label{Sec:Conclusion}

\vspace*{1mm}
\noindent
We have calculated the gluonic contributions to the massive form factors in the vector, 
axial-vector, scalar and 
pseudoscalar cases, adding on the quarkonic contributions~\cite{Blumlein:2023uuq} computed by us previously.
These are all non-singlet contributions. We have performed these calculations as widely as possible 
analytically. The computer-algebraic efforts to perform the present calculation were extremely large, starting
with the integration-by-parts reduction, the calculation of
deep analytic expansions in the low energy limit by using the method of arbitrary high moments 
\cite{Blumlein:2017dxp}, and
the guessing of the necessary difference and differential equations. The matching between
the different points in $s$ maintaining very high accuracy constitutes a further
challenge, to cover the complete kinematic region analytically in terms of very deep 
expansions. The guessing of the initial difference and differential equations at 
$s = 0$ required massive parallelization in the most complicated cases and 
the use of large mainframes such as the super-computer {\tt MACH-2}. 
The differential and difference equations obtained in this way belong to the most 
voluminous ones having been computed so far.

The present  calculation has been possible by using very deep Frobenius-expansions of the 
contributing 
differential equations, which are used mapping out the whole kinematic region at high numerical accuracy.
Starting from $s=0$ we reached in this way an analytic expansion around 
$s \rightarrow  \infty$ for the first time 
and determined all contributing constants by using integer algorithms. Contrary to the quarkonic case, also
non multiple zeta values contribute, which finally are found to belong to the class of iterated integrals described 
by some of us in Refs.~\cite{Ablinger:2011te,Ablinger:2021fnc}. The $s$-dependence of some of the contributions
could even be determined in closed form in terms of iterated integrals.

We have performed detailed comparisons with the numerical results presented in 
Refs.~\cite{Fael:2022miw,Fael:2022rgm,KIT1,MATKIT} before and found relative agreement 
of better than $O(10^{-10})$.
In ancillary files to this paper we present computer-readable analytic results. Numerical 
codes are also available.

\appendix 
\section{Codes}
\label{sec:A}

\vspace*{1mm}
\noindent
The following {\tt Mathematica} codes are provided for the gluonic 
contributions to the 
form factors in form of ancillary files. The notebook {\tt EvaluateNumerical.nb} 
evaluates the 
unrenormalized $O((\alpha_s/\pi)^3)$ part of the formfactors to $O(10^{-30})$ accuracy in the whole 
kinematic
range of $s$. The file system is contained in {\tt t1.tar.gz}.
Here the analytic pieces of the type (\ref{eq:solvable-fv1}, \ref{eq:solvable-fv2}) 
are not included. They are given as {\tt .m} files in the file {\tt t0.tar.gz}.
The guessed color-zeta factors as functions, cf.~(\ref{asyCF3V1}), for the cases 
V1 and V2 for all color factors are given in the file {\tt t2.tar.gz}.
The numerical representations of the six unrenormalized and renormalized gluonic form 
factors to $O(a_s^3)$
are available by the code {\tt FFNUM2.f}. It is available on request via e-mail from
{\tt Johannes.Bluemlein@desy.de}.

\vspace*{5mm}
\noindent
{\bf Acknowledgment.}
We thank DI J.~Messner from Zentraler Informationsdienst (ZID) Johannes Kepler
University, Linz, for installing the parallelized guessing software \cite{GSAGE} 
at the supercomputer {\tt MACH-2} of JKU Linz and his help in maintaining the runs
determining the largest differential and difference operators in this project.
We thank M.~Kauers and K.~Sch\"onwald for discussions, and D.~Broadhurst for
his interest in our work. This research was funded in whole or in part by the 
Austrian Science Fund (FWF) grants DOI 10.55776/P33530,
DOI 10.55776/PAT1332123, and DOI 10.55776/I6130.

\vspace{5mm}
{\footnotesize

}
\end{document}